\documentclass[a4paper,11pt]{article}

\usepackage{jheppub}

\usepackage{amsfonts,graphics,epsfig,subfigure}

\title{A unified phase transition picture of the charged topological black hole in Ho\v{r}ava-Lifshitz gravity}

\author[a,b]{Jie-Xiong Mo\note{mojiexiong@gmail.com},}
\author[c,a]{Xiao-Xiong Zeng,}
\author[b]{Gu-Qiang Li,}
\author[a]{Xin Jiang,}
\author[a,1]{Wen-Biao Liu\note{Corresponding
author:wbliu@bnu.edu.cn}}

\affiliation[a]{Department of Physics, Institute of Theoretical
Physics, Beijing Normal University,\\Beijing, 100875, China}
\affiliation[b]{Institute of Theoretical Physics, Zhanjiang Normal
University,\\Zhanjiang,Guangdong, 524048, China}
\affiliation[c]{School of Science, Chongqing Jiaotong University,\\
Nanan,Chongqing, 400074, China}

\abstract{Aiming at a unified phase transition picture of the
charged topological black hole in Ho\v{r}ava-Lifshitz gravity, we
investigate this issue not only in canonical ensemble with the fixed
charge case but also in grand-canonical ensemble with the fixed
potential case. We firstly perform the standard analysis of the
specific heat, the free energy and the Gibbs potential, and then
study its geometrothermodynamics. It is shown that the local phase
transition points not only witness the divergence of the specific
heat, but also witness the minimum temperature and the maximum free
energy or Gibbs potential. They also witness
the divergence of the corresponding thermodynamic scalar curvature.
No matter which ensemble is chosen, the metric constructed can successfully
produce the behavior of the thermodynamic interaction and phase
transition structure while other metrics failed to predict the phase
transition point of the charged topological black hole in former literature. In grand-canonical ensemble, we have discovered the
phase transition which has not been reported before. It is similar to the canonical ensemble in
which the phase transition only takes place when $k=-1$. But it also
has its unique characteristics that the location of the phase
transition point depends on the value of potential, which is
different from the canonical ensemble where
the phase transition point is independent of the parameters. After
an analytical check of Ehrenfest scheme, we find that the new phase
transition is a second order one. It is also found that the
thermodynamics of the black hole in Horava-Lifshitz gravity is quite
different from that in Einstein gravity.}

\begin{document}
\maketitle
\flushbottom

\section{Introduction}
\label{Sec1}
    Black hole thermodynamics has become a fascinating topic in the theoretical physics since Bekenstein and
     Hawking investigated the black hole entropy and identified black holes as thermodynamic objects~\cite{Bekenstein}, \cite{Hawking1}.Various thermodynamic
     properties of black holes have been studied widely including the phase transition. In 1983, Hawking and
     Page discovered that there exists a phase transition between the Schwarzschild AdS black hole and the thermal AdS space~\cite{Hawking2}.
     From then on, phase transitions of black holes have aroused theoretical physicists' attention and have been intensely investigated from
     different perspectives~\cite{Chamblin1}-\cite{Eune}.

     On the one hand, traditional thermodynamics has been applied to investigate the critical behavior of a black hole.
For example, utilizing Clausius-Clapeyron-Ehrenfest¡¯s equations,
one can classify phase transitions as first order or higher order
transitions. For a first order transition the Clausius-Clapeyron
equation is satisfied. While for a second order transition
Ehrenfest¡¯s relations are satisfied. Recently, Banerjee et al.
developed a scheme based on Ehrenfest¡¯s relations to study phase
transitions in black holes~\cite{Banerjee1}-\cite{Banerjee6}. They
considered the black holes as grand-canonical ensembles and
performed a detailed analysis of Ehrenfest¡¯s relations using both
analytical and graphical techniques.

    On the other hand, geometric approach has served as an alternative way to study phase transitions of black holes.
    Riemannian geometry in the space of equilibrium states was introduced by Weinhold~\cite{Weinhold} and Ruppeiner~\cite{Ruppeiner}.
    Weinhold proposed metric structure in the energy representation as $g_{i,j}^{W}=\partial_{i}\partial_{j}M(U,N^a)$, which was the Hessian
    matrix of the internal energy
    $U$
    with respect to the extensive thermodynamic variables $N^a$. Ruppeiner defined metric structure as $g_{i,j}^{R}=-\partial_{i}\partial_{j}S(U,N^a)$, which was the Hessian of the entropy.
    However, Weinhold's and Ruppeiner's metrics are not invariant under Legendre transformations and sometimes lead to contradictory results~\cite{Salamon}-\cite{Mrugala}.
    Taking Legendre invariance into consideration, Quevedo et al. \cite{Quevedo2} recently presented a new formalism of geometrothermodynamics, which
    allows us to derive Legendre invariant metrics in the space of equilibrium states. Geometrothermodynamics  presents a unified geometry
    where the metric structure can give a well description of various types of black hole thermodynamics ~\cite{Quevedo3}-\cite{Hanyiwen}.

    Here, we would like to focus our attention on the phase transition of the charged topological black hole in Ho\v{r}ava-Lifshitz gravity.
    Ho\v{r}ava-Lifshitz (HL) theory is a non-relativistic renormalizable theory of gravity at a Lifshitz point proposed by Ho\v{r}ava~\cite{Horava1}-\cite{Horava3}.
    HL gravity provides a fascinating framework for one to explore the connections between ordinary gravity and string theory.
    The black hole solutions \cite{Luhong}-\cite{Kiritsis} and thermodynamic properties~\cite{Cai5}-\cite{Myung5} have attracted a lot of attention.
    Concerning the phase transition, some efforts have also been made. Koutsoumbas et al.~\cite{Koutsoumbas} mainly discussed the perturbative behaviour
    and quasi-normal modes of charged topological AdS black holes. Cao et al.~\cite{Cao} studied black hole phase transitions in (deformed) HL gravity,
    including the charged/uncharged topological black holes and Kehagias-Sfetsos (KS) black hole. However, their geometric approach was based on
    the Ruppeiner and Weinhold metrics and failed to predict the phase transition. Quevedo et al.~\cite{Quevedo1}-\cite{Quevedo8} investigated the geometrothermodynamics in
    HL gravity. But it mainly handled the Cai-Cao-Ohta (CCO) topological black holes and left the charged topological black holes uninvestigated. Wei et al.~\cite{Weishaowen1} mainly
discussed the thermodynamic geometry and phase transition of KS
black hole in the deformed HL gravity while Majhi et
al.~\cite{Majhi} focused their attention
    on the scaling behavior of topological charged black holes in HL gravity. In this paper, we would like to further elaborate the research on
    the phase transition of the charged topological black hole in HL gravity. Aiming at a unified picture, the phase transition would
    be considered not only in canonical ensemble with the fixed charge case but also in grand-canonical ensemble with the fixed
    potential case. Both geometrothermodynamics and the Ehrenfest¡¯s scheme would be applied to carry out the research.

    The organization of our paper is as follows. In Section~\ref{Sec2}, the thermodynamics of the charged topological
black hole in HL gravity will be reviewed briefly. In Section~\ref {Sec3}, the charged
topological black hole as canonical ensemble will be investigated in
geometrothermodynamics. In Section~\ref {Sec4}, the phase transition
will be studied taking the black hole as grand-canonical ensemble.
To investigate the nature of the new phase transition in
Section~\ref {Sec4}, an analytical check of Ehrenfest equations will
be carried out in Section~\ref {Sec5}. In the end, a discussion is
given in Section~\ref {Sec6}.

\section{Review of thermodynamics of the charged topological black hole}
\label{Sec2} The charged topological black hole solution in
Ho\v{r}ava-Lifshitz gravity has been discussed in~\cite{Cai4}. For
simplicity, the dynamical coupling constant $\lambda$ can be set to
one. And the metric is given as
\begin{align}
ds^2&=-\widetilde{N}(r)^2dt^2+\frac{dr^2}{f(r)}+r^2d\Omega_k^2,\label{1}\\
f(r)&=k+x^2-\sqrt{c_0x-\frac{q^2}{2}},\qquad
x=\sqrt{-\Lambda}r,\label{2}
\end{align}
where $\Lambda $ corresponds to the negative cosmological constant
and $d\Omega_k$ is the line element of a two dimensional Einstein
space with constant scalar curvature 2$k$. Without loss of
generality, one can take $k = 0, \pm1$ respectively. For the metric
given above, $\widetilde{N}=N_0$ could be set to one. Solving the
equation $f(r) =0$, we can get the largest positive root, from which
we can determine the event horizon radius. Denoting
$l^2=-\frac{1}{\Lambda}$, the relevant quantities have been reviewed
in~\cite{Cao} as
\begin{align}
T&=\frac{6x_+^4+4kx_+^2-2k^2-q^2}{16kl^2\pi x_++16l^2\pi
x_+^3},\label{3}\\
S&=\frac{\pi\kappa^2\mu^2\Omega_k}{4}(x_+^2+2k\ln(x_+))+S_0,\label{4}\\
\Phi&=\frac{q}{x_+}+\Phi_0,\label{5}\\
Q&=\frac{\kappa^2\mu^2\Omega_k}{16l^2}q,\label{6}\\
M&=\frac{\kappa^2\mu^2\Omega_k}{16l^2}c_0,\label{7}
\end{align}
where $c_0=\frac{2k^2+q^2+4kx_+^2+2x_+^4}{2x_+},  c_0,q$ are the
integration constants and $\kappa,\mu $ are the constant parameters
of the theory. $\Omega_k$ is the volume of two dimensional Einstein
space.
\begin{figure*}
\centerline{\subfigure[]{\label{1a}
\includegraphics[width=8cm,height=6cm]{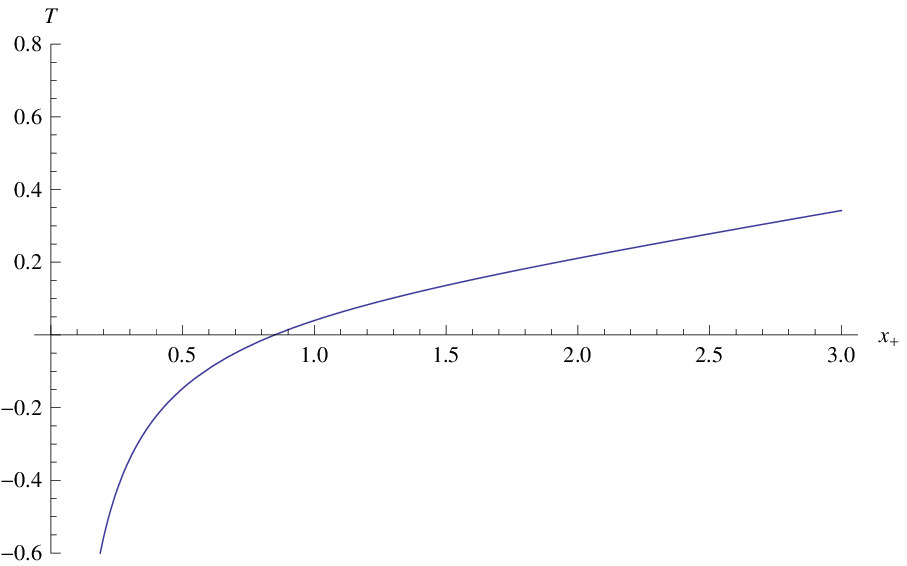}}
\subfigure[]{\label{1b}
\includegraphics[width=8cm,height=6cm]{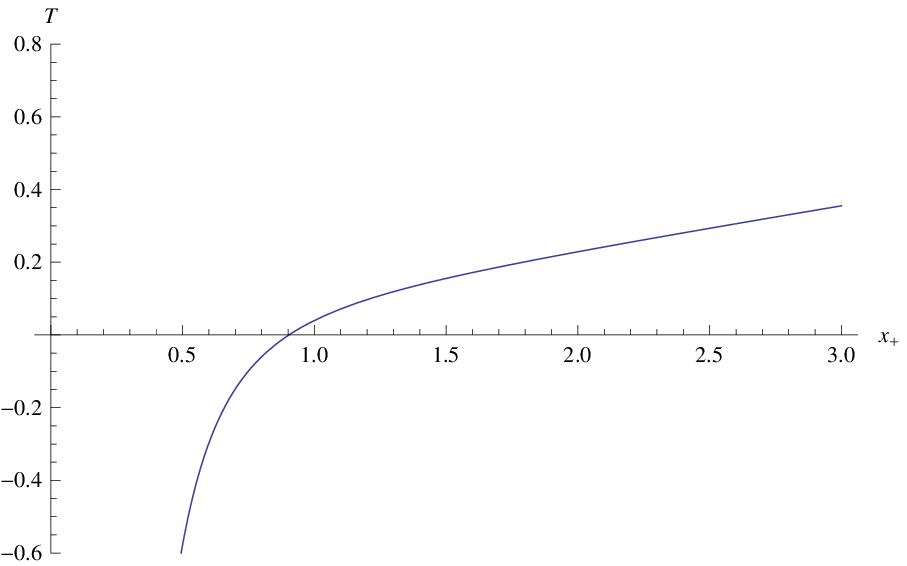}}}
\subfigure[]{\label{1c}
\includegraphics[width=8cm,height=6cm]{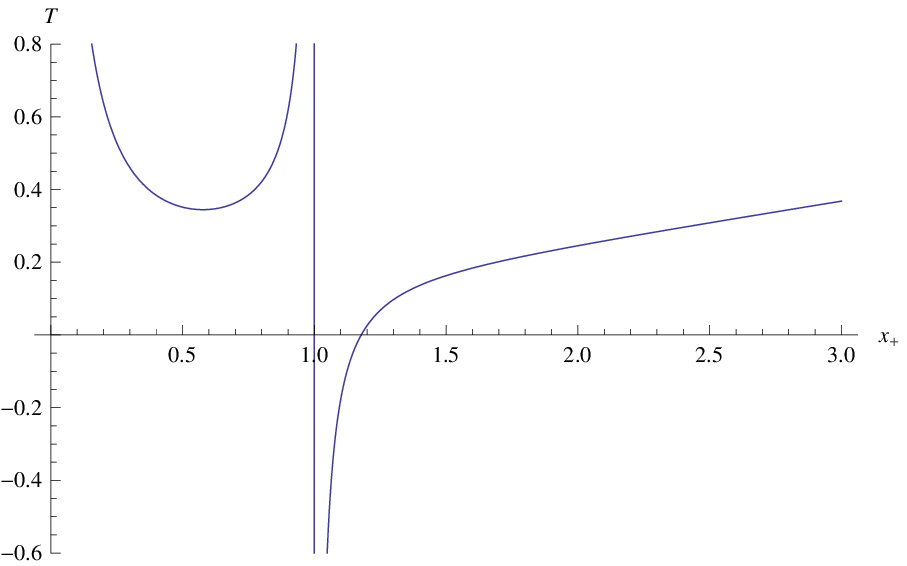}}
 \caption{Hawking temperature $T$ vs. $x_+$ for
(a) $k=1,q=2$ (b) $k=0,q=2$ (c) $k=-1,q=2$} \label{fg1}
\end{figure*}

    From Eq.(\ref{3}), we can easily find that the charged topological black hole becomes an extremal one when $q^2=6x_+^4+4kx_+^2-2k^2$.
    Hawking temperature will be negative when $6x_+^4+4kx_+^2-2k^2<q^2$ , which implies the existence of some unphysical regions.
    To show the variation of Hawking temperature explicitly, we plot Figure~\ref{fg1} using Eq.(\ref{3}).
    In Figure~\ref{fg1}, we exhibit the temperature $T$ vs. $x_+$ respectively for the cases $k = 0,
    \pm1$
    (Note that parameters are chosen as $c = l = G =\kappa=\mu=\Omega_k= 1$ in all figures in this paper.)

        From Figure~\ref{fg1}, we can find that each case has some unphysical regions with negative Hawking temperature. For the case $k = 0,1$, Hawking temperature increases
         monotonically. However, things are different for the case $k=-1$. $x_+=1$ divides the region into two parts.
        When $x_+>1$, Hawking temperature increases monotonically. When $0<x_+<1$, there exists a minimum
        Hawking temperature, which can be derived from
\begin{equation}
\frac{\partial T_{k=-1}}{\partial x_+}=\frac{\partial}{\partial x_+}
(\frac{6x_+^4-4x_+^2-2-q^2}{-16l^2\pi x_++16l^2\pi
x_+^3})=0.\label{8}
\end{equation}%

    Solving Eq.(\ref{8}), we can get the corresponding $x_+$ and the minimum Hawking temperature as
 \begin{equation}
x_+=\frac{\sqrt{3}}{3},\quad T_{min}=\frac{\sqrt{3}(8+3q^2)}{32\pi
l^2}.\label{9}
\end{equation}%
Note that the location of $x_+$ that corresponds to the minimum
Hawking temperature is independent of the charge parameter $q$.

\section{Phase transition and geometrothermodynamics in fixed-charge ensemble}
\label{Sec3}
    When the charge of black hole is fixed, the specific heat can be given as
 \begin{equation}
C_Q=T(\frac{\partial S}{\partial T})_Q=\frac{-\pi
(k+x_+^2)^2[q^2+2(k-3x_+^2)(k+x_+^2)]\kappa^2\Omega_k\mu^2}{2(k+3x_+^2)[q^2+2(k+x_+^2)^2]}.\label{10}
\end{equation}%

    From Eq.(\ref{10}), we can conclude that $C_Q$ diverges when $k+3x_+^2=0$ . The equation $k+3x_+^2=0$ has positive root only when $k=-1$. And the root
    can be solved as $x_+=\frac{\sqrt{3}}{3}$. It is quite interesting to note that the point where $C_Q$ diverges is independent of charge
    parameter $q$ and coincides with the point corresponding to the minimum Hawking temperature.

\begin{figure*}
\centerline{\subfigure[]{\label{2a}
\includegraphics[width=8cm,height=6cm]{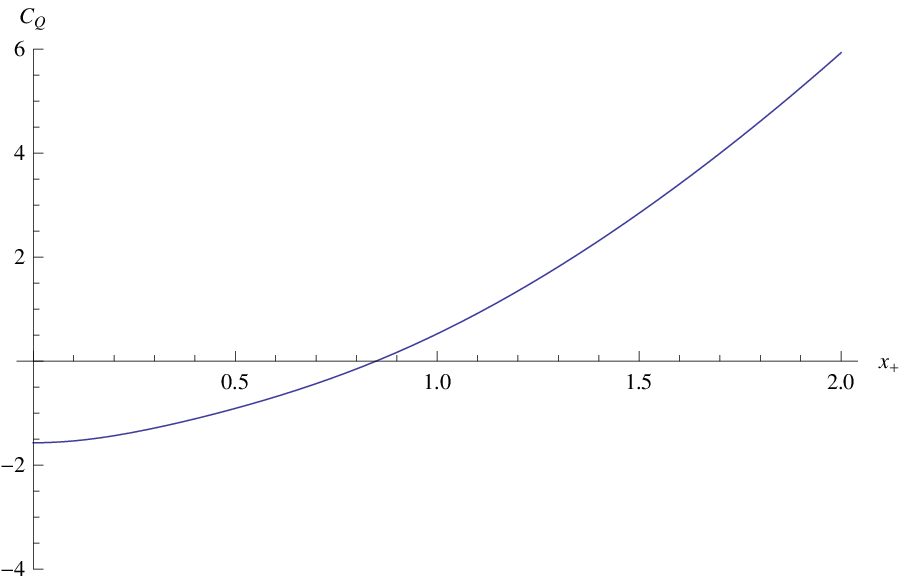}}
\subfigure[]{\label{2b}
\includegraphics[width=8cm,height=6cm]{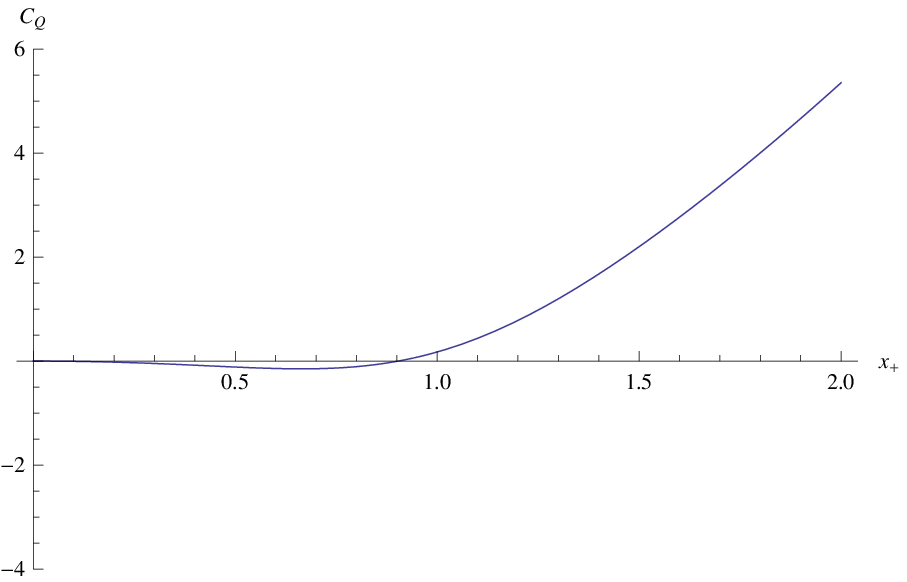}}}
\subfigure[]{\label{2c}
\includegraphics[width=8cm,height=6cm]{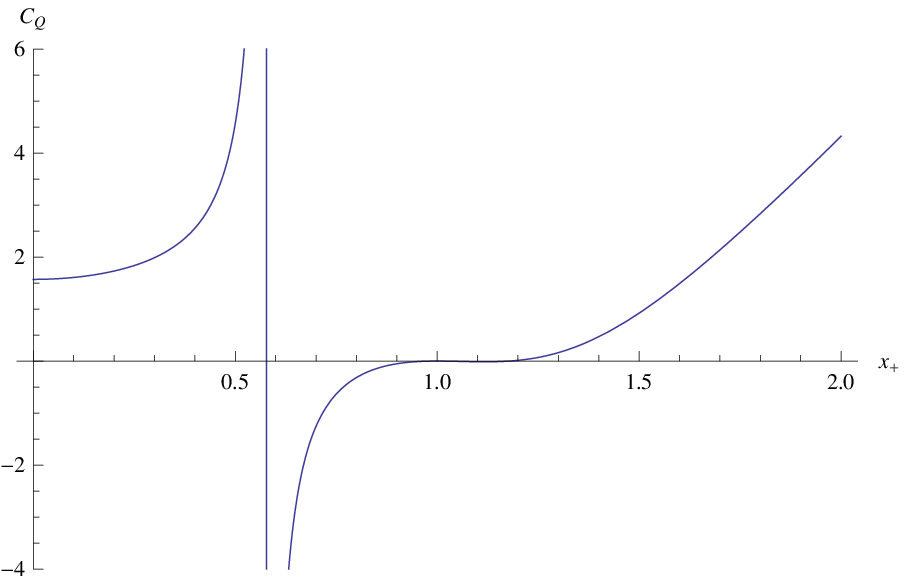}}
 \caption{$C_Q$ vs. $x_+$ for
(a) $k=1,q=2$ (b) $k=0,q=2$ (c) $k=-1,q=2$} \label{fg2}
\end{figure*}

        To observe the possible divergence of $C_Q$, we plot Figure~\ref{fg2} using Eq.(\ref{10}). In Figure~\ref{fg2}, we exhibit the behavior
        of $C_Q$ respectively for the cases $k = 0, \pm1$. We can see that $C_Q$ is continuous for the
cases $k = 0, 1$ and no phase transition takes place. However, the
curve for the case $k=-1$ gives an infinite discontinuity, which
suggests the existence of phase transition. The phase transition
location is $x_+=\frac{\sqrt{3}}{3}$.

    To confirm the phase transition in canonical ensemble, we would like to perform an
    analysis of the behavior of free energy. The free energy which is defined
    by $F=M-TS$ can be obtained as
 \begin{align}
F=&\frac{\kappa^2\mu^2\Omega_k}{64l^2x_+(k+x_+^2)}\times[4k^3+14k^2x_+^2+8kx_+^4-2x_+^6+2kq^2+3q^2x_+^2\nonumber
\\
&+kln x_+(4k^2+2q^2-8kx_+^2-12x_+^4)]
 .\label{99}
\end{align}
\begin{figure*}
\centerline{\subfigure[]{\label{3a}
\includegraphics[width=8cm,height=6cm]{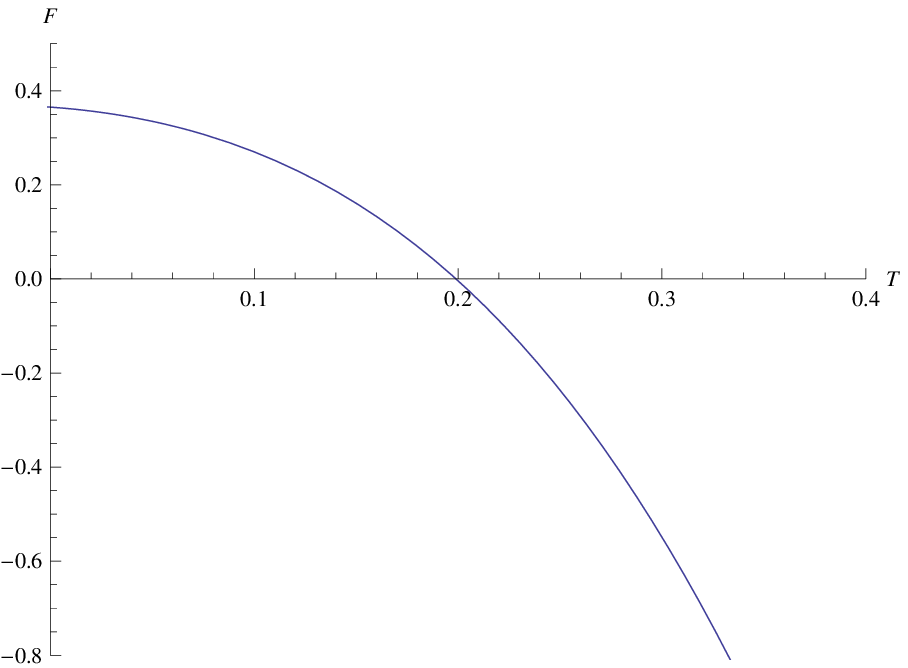}}
\subfigure[]{\label{3b}
\includegraphics[width=8cm,height=6cm]{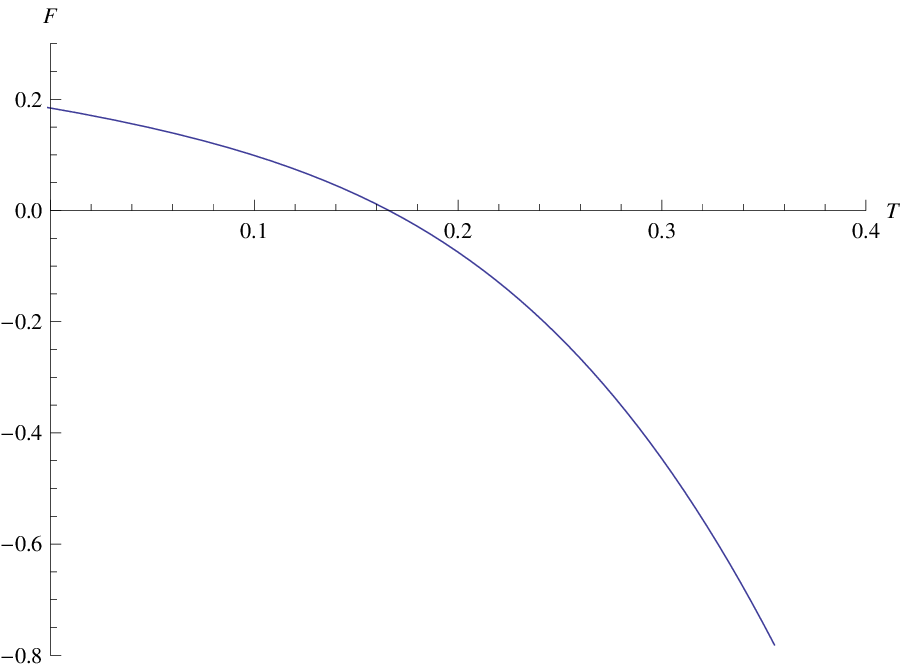}}}
\subfigure[]{\label{3c}
\includegraphics[width=8cm,height=6cm]{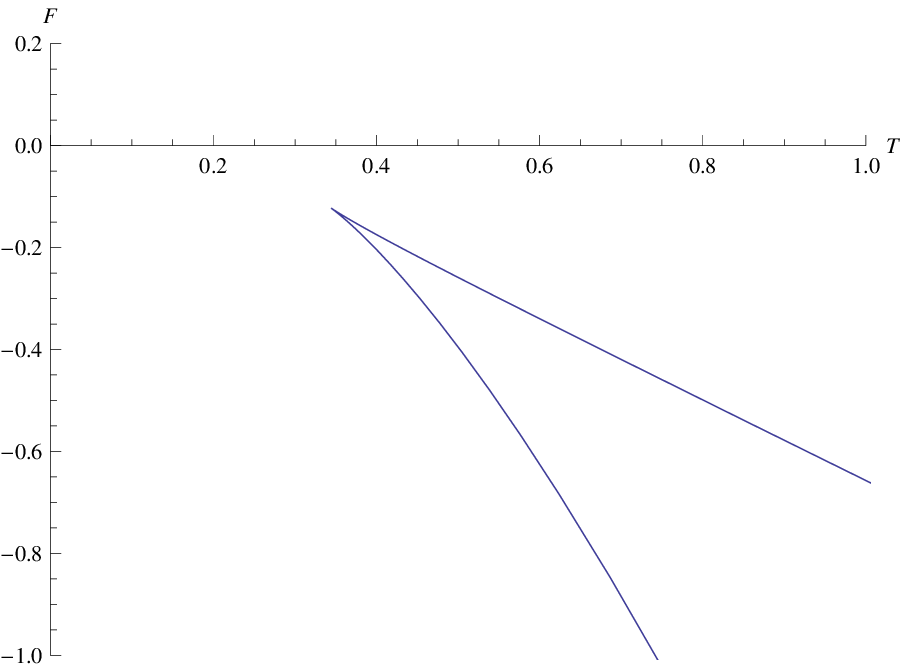}}
 \caption{$F$ vs. $T$ for
(a) $k=1,q=2$ (b) $k=0,q=2$ (c) $k=-1,q=2$} \label{fg3}
\end{figure*}
Figure~\ref{fg3} shows the free energy vs. the
temperature for three different cases. For $k=1,0$, the free energy
decreases steadily when the temperature increases. However, for
$k=-1$, two different phases are described by two wings which are
joined at the point where the free energy reaches a maximum value.
According to traditional thermodynamics, the system is most unstable
at this point and would eventually passes to the locally stable
phase which corresponds to the lower wing. It is quite interesting
to note that the point of the maximum free energy also
corresponds to the minimum Hawking temperature. As we stated before,
it is also the point where the local phase transition takes place
from a locally unstable small black hole to a locally stable large
black hole. So we can conclude that the local phase transition
points not only witness the divergence of the specific heat but also
witness the minimum temperature and maximum free energy.

    Before we go further to investigate its geometrothermodynamics, we would like to briefly review the construction of geometrothermodynamics.
    According to Ref.~\cite{Quevedo2}, the $(2n+1)$-dimensional thermodynamic phase space $\mathcal {T}$ can be coordinated by the set of independent quantities \{$\phi,E^a,I^a$\},
     where $a = 1, \cdots, n$, the positive integer $n$ represents the number of thermodynamic degree of freedom of the system, $\phi$ corresponds to
     the thermodynamic potential, and $E^a,I^a$ are the extensive and intensive thermodynamic variables respectively. The fundamental Gibbs 1- form defined on $\mathcal {T}$ can then be written as $\Theta=d\phi-\delta_{ab}I^adE^b$, where $\delta_{ab}=diag(1,\cdots,1)$.
      Considering a non-degenerate Riemannian metric $G$, a contact Riemannian manifold can be defined from the set $(\mathcal {T},\Theta,G)$ if the
      condition $\Theta\wedge(\Theta)^n\neq0$ is satisfied. Legendre transformation is such a special case of contact transformation that
      it can keep the contact structure of $\mathcal {T}$ invariant. Legendre invariance guarantees that the geometric properties of $G$ do not depend on
      the thermodynamic potential. Utilizing a smooth map $\varphi:\varepsilon\rightarrow\mathcal{T}$, i.e. $\varphi:(E^a)\mapsto(\phi,E^a,I^a)$,
      a submanifold $\varepsilon$ called the space of thermodynamic equilibrium states can be induced. $\phi=\phi(E^a)$ and the map satisfies the
      condition $\varphi^*(\Theta)=\varphi^*(d\phi-\delta_{ab}I^adE^b)=0$, where $\varphi^*$ is the pullback of $\varphi$. $\varepsilon\subset\mathcal {T}$
      and $\varepsilon$ is an $n$-dimensional Riemannian submanifold. Furthermore, a thermodynamic metric $g$ can be induced in a canonical manner in
      the equilibrium manifold $\varepsilon$ by the smooth map $\varphi$.

          As proposed by Quevedo, the non-degenerate metric $G$ and the thermodynamic metric $g$ can be written as
          follows~\cite{Quevedo7}
\begin{align}
G&=(d\phi-\delta_{ab}I^adE^b)^2+(\delta_{ab}E^aI^b)(\eta_{cd}dE^cdI^d),\label{11}\\
g&=\varphi^*(G)=(E^c\frac{\partial \phi}{\partial
E^c})(\eta_{ab}\delta^{bc}\frac{\partial^2\phi}{\partial E^c
\partial E^d}dE^adE^d),\label{12}
\end{align}%
where $\eta_{ab}=diag(-1,\cdots,1)$.

    Now let$\textquoteright$s begin to apply geometrothermodynamics to investigate the phase
transition of the charged topological
    black hole in HL gravity. To construct geometrothermodynamics in fixed-charge ensemble, we choose $M$ as the thermodynamic potential,
    and $S,Q$ as extensive variables. The corresponding thermodynamic phase space is a 5-dimensional one coordinated by the set of
    independent coordinates \{$M,S,Q,T,\Phi$\}. The fundamental Gibbs 1- form defined on $\mathcal {T}$ can be written as
\begin{equation}
\Theta=dM-TdS-\Phi dQ,\label{13}
\end{equation}%
and the non-degenerate metric $G$ from Eq.(\ref{11}) is
\begin{equation}
G=(dM-TdS-\Phi dQ)^2+(TS+\Phi Q)(-dSdT+dQd\Phi).\label{14}
\end{equation}%
Introducing the map
  \begin{equation}
\varphi:\{S,Q\}\mapsto\{M(S,Q),S,Q,\frac{\partial M}{\partial
S},\frac{\partial M}{\partial Q}\},\label{15}
\end{equation}%
the space of thermodynamic equilibrium states can be induced.
According to Eq.(\ref{12}), the thermodynamic metric $g$ can be
written as follows
\begin{equation}
g=(S\frac{\partial M}{\partial S}+Q\frac{\partial M}{\partial
Q})(-\frac{\partial^2M}{\partial S^2}dS^2+\frac{\partial^2
M}{\partial Q^2}dQ^2).\label{16}
\end{equation}%
Utilizing Eqs.(\ref{4}),(\ref{6})and (\ref{7}), we can easily
calculate the relevant quantities in Eq.(\ref{16}) as below
\begin{align}
\frac{\partial M}{\partial
S}&=\frac{-q^2-2(k-3x_+^2)(k+x_+^2)}{16l^2\pi
x_+(k+x_+^2)},\label{17}\\
\frac{\partial M}{\partial Q}&=\frac{q}{x_+},\label{18}\\
\frac{\partial^2 M}{\partial
S^2}&=\frac{(k+3x_+^2)[q^2+2(k+x_+^2)^2]}{8l^2\pi^2
x_+(k+x_+^2)^2\kappa^2\mu^2\Omega_k},\label{19}\\
\frac{\partial^2 M}{\partial
Q^2}&=\frac{16l^2}{\kappa^2\mu^2\Omega_kx_+}.\label{20}
\end{align}%
Comparing Eqs.(\ref{17}),(\ref{18}) with Eqs.(\ref{3}),(\ref{5}), we
find
\begin{equation}
\frac{\partial M}{\partial S}=T,\quad \frac{\partial M}{\partial
Q}=\Phi-\Phi_0.\label{21}
\end{equation}%

\begin{figure*}
\centerline{\subfigure[]{\label{4a}
\includegraphics[width=8cm,height=6cm]{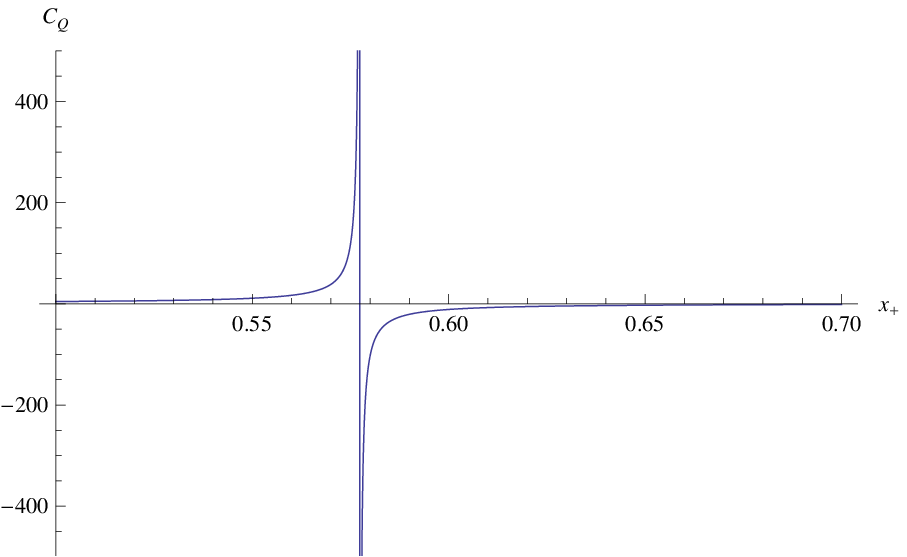}}
\subfigure[]{\label{4b}
\includegraphics[width=8cm,height=6cm]{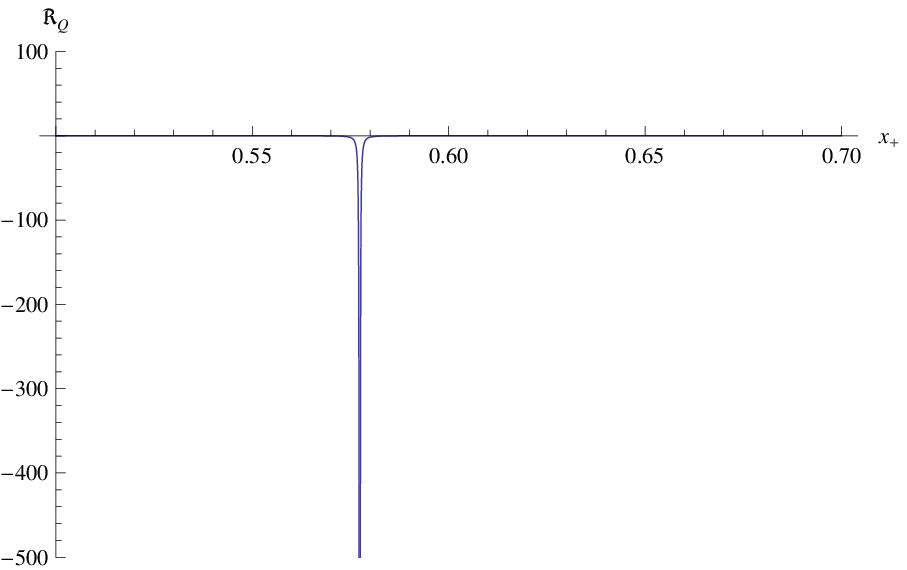}}}
 \caption{($a$) $C_Q$ vs. $x_+$ for $k=-1,Q=2$ ($b$) $\mathfrak{R}_Q$ vs. $x_+$ for $k=-1,Q=2$ }
\label{fg4}
\end{figure*}

 To assure that the first law of black hole
thermodynamics holds,we can derive from Eq.(\ref{21}) $\Phi_0=0$.
Substituting Eqs.(\ref{17})-(\ref{20}) into Eq.(\ref{16}), we can
calculate the component of the thermodynamic metric $g$ as
\begin{align}
g_{QQ}&=\frac{q^2}{x_+^2}-\frac{[q^2+2(k+x_+^2)(k-3x_+^2)][4S_0+\pi
\kappa^2\mu^2\Omega_k(x_+^2+2k\ln x_+)]}{4\pi
x_+^2(k+x_+^2)\kappa^2\mu^2\Omega_k},\label{22}\\
g_{SS}&=-\frac{A(x_+,q)}{512l^4\pi^3
\kappa^2\mu^2\Omega_kx_+^2(k+x_+^2)^4},\label{23}
\end{align}%
where
\begin{align}
A(x_+,q)=&(k+3x_+^2)[q^2+2(k+x_+^2)^2]\nonumber
\\
&\times\{-[q^2+2(k-3x_+^2)(k+x_+^2)][4S_0+\pi\kappa^2\mu^2\Omega_k(x_+^2+2k\ln
x_+)]\nonumber
\\
&+4q^2\pi\kappa^2\mu^2\Omega_k (k+x_+^2)\}. \label{24}
\end{align}%
Till now, we can obtain the Legendre invariant scalar curvature as
\begin{equation}
\mathfrak{R}_Q=\frac{B(x_+,Q)}{D(x_+,Q)},\label{25}
\end{equation}%
where
\begin{align}
D(x_+,Q)=&(k+3x_+^2)^2\times(128l^4Q^2+\kappa^4\mu^4\Omega_k^2(k+x_+^2)^2)^2
\nonumber
\\
&\times\{512S_0l^4Q^2-128\pi l^4Q^2\kappa^2\mu^2\Omega_k(4k+3x_+^2)
\nonumber
\\
&+4S_0\kappa^4\mu^4\Omega_k^2(k-3x_+^2)(k+x_+^2)+\pi
\kappa^6\mu^6\Omega_k^3x_+^2(k-3x_+^2)(k+x_+^2) \nonumber
\\
&+2k\pi\kappa^2\mu^2\Omega_k[128l^4Q^2+\kappa^4\mu^4\Omega_k^2(k-3x_+^2)(k+x_+^2)]\ln
x_+\}^3. \label{26}
\end{align}%
From Eq.(\ref{26}), we can find that the Legendre invariant scalar
curvature diverges when $k+3x_+^2=0$, which corresponds to
$k=-1,x_+=\frac{\sqrt{3}}{3}$. That is the exact point where the
phase transition takes place. To get an intuitive sense on this
issue, we plot Figure~\ref{fg4}, which shows the correpondence of
the divergence of specific heat $C_Q$ and the thermodynamic scalar
curvature of $\mathfrak{R}_Q$. From Figure~\ref{fg4}, we learn that
the Legendre invariant metric constructed in geometrothermodynamics
correctly produces the behavior of the thermodynamic interaction and
phase transition structure.

\section{Phase transition and geometrothermodynamics in fixed-potential ensemble}
\label{Sec4}
\begin{figure*}
\centerline{\subfigure[]{\label{5a}
\includegraphics[width=8cm,height=6cm]{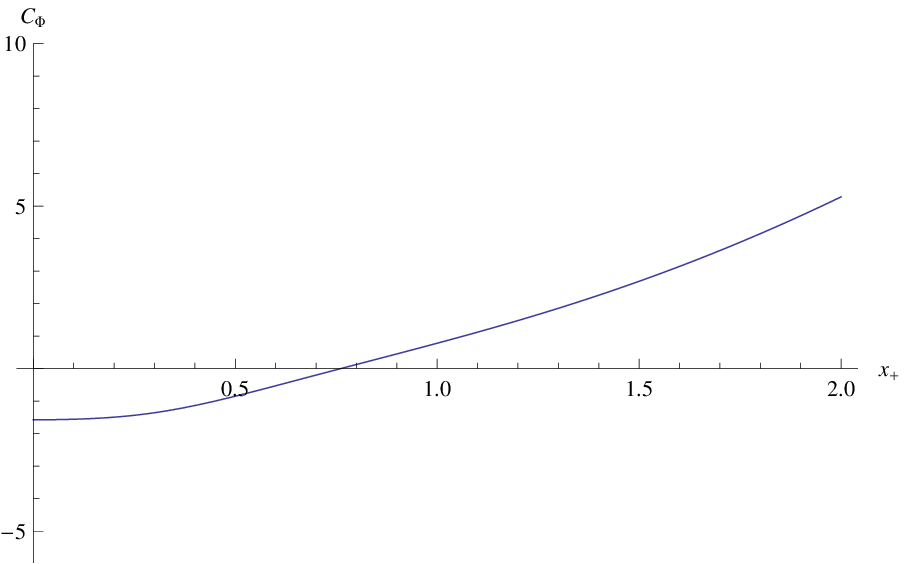}}
\subfigure[]{\label{5b}
\includegraphics[width=8cm,height=6cm]{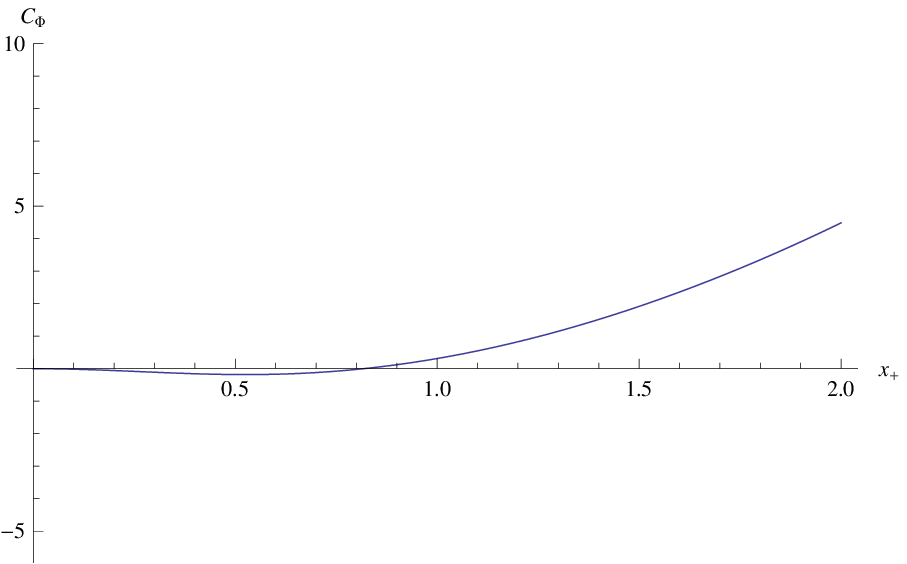}}}
\centerline{\subfigure[]{\label{5c}
\includegraphics[width=8cm,height=6cm]{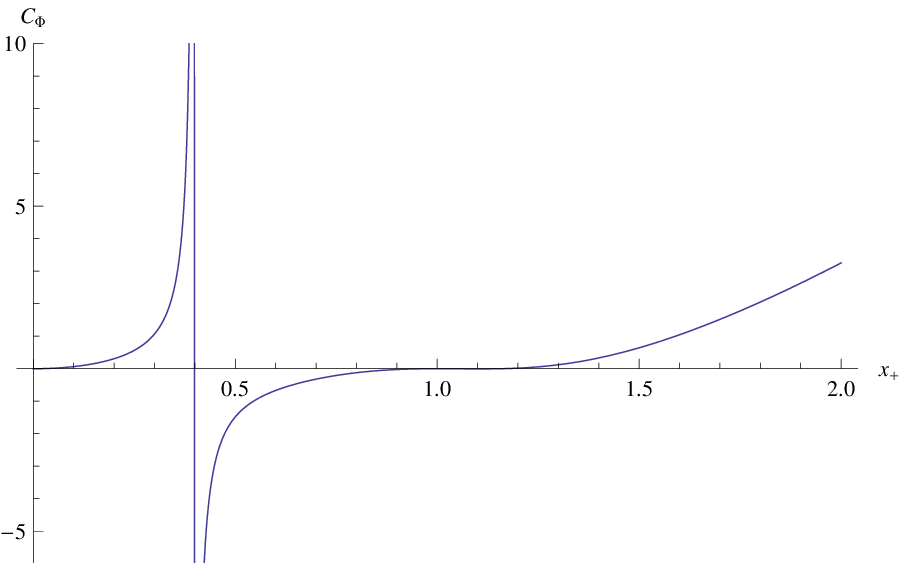}}}
 \caption{$C_\Phi$ vs. $x_+$ for
(a) $k=1,\Phi=2$ (b) $k=0,\Phi=2$ (c) $k=-1,\Phi=2$} \label{fg5}
\end{figure*}

When the potential of black hole is fixed, the
specific heat is shown as
\begin{equation}
C_\Phi=T(\frac{\partial S}{\partial T})_\Phi=\frac{-\pi
\kappa^2\mu^2\Omega_k
(k+x_+^2)^2(2k^2-4kx_+^2-6x_+^4+x_+^2\Phi^2)}{2(2k^3+10k^2x_+^2+14kx_+^4-kx_+^2\Phi^2+x_+^4\Phi^2+6x_+^6)}.
\label{27}
\end{equation}%
Apparently, $C_\Phi$ may diverge when
\begin{equation}
2k^3+10k^2x_+^2+14kx_+^4-kx_+^2\Phi^2+x_+^4\Phi^2+6x_+^6=0,
\label{28}
\end{equation}%
which suggests a possible phase transition. However, the phase
transition point characterized by Eq.(\ref{28}) is not intuitive.
    To gain an intuitive understanding, we plot Figure~\ref{fg5} using Eq.(\ref{27}) . In Figure~\ref{fg5},
    we exhibit the behavior of $C_\Phi$ respectively for the cases $k = 0, \pm1$.
We can find that $C_\Phi$ diverges only for
$k=-1$, which is similar to the fixed-charged ensemble. To check
whether the phase transition point locates in the physical region,
 we also plot the Hawking temperature $T$ vs. $x_+$ for the case $k=-1$ in Figure~\ref{fg5}(c). It is shown that the phase
 transition point locates in the positive temperature region. Figure~\ref{fg5}(c)
can be divided into two phases. One is thermodynamically stable
($C_\Phi>0$) with small radius while the other is unstable
($C_\Phi<0$) with large radius. So the phase transition takes place
between small black hole and large black hole.

\begin{figure*}
\centerline{\subfigure[]{\label{6a}
\includegraphics[width=8cm,height=6cm]{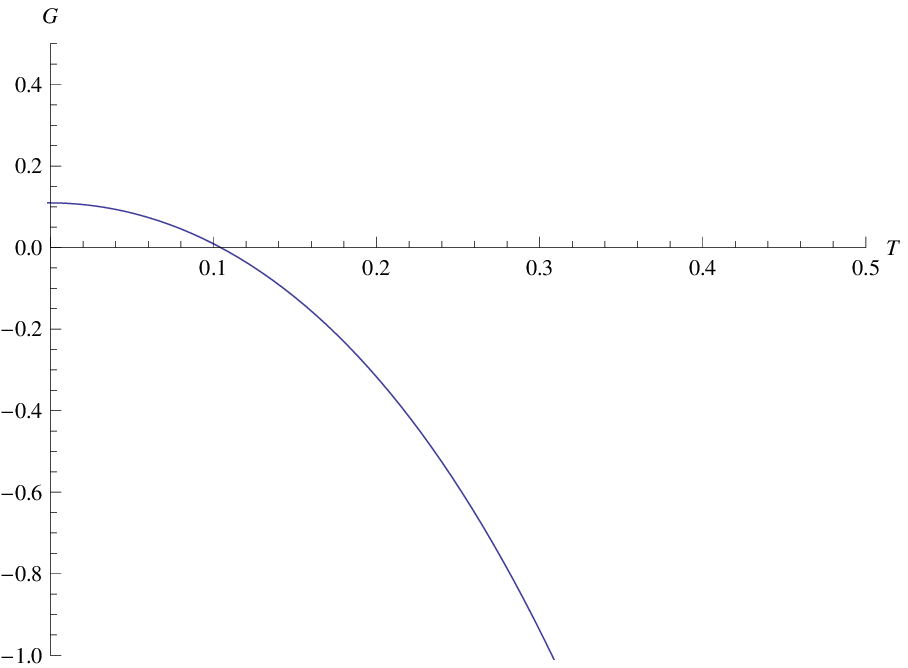}}
\subfigure[]{\label{6b}
\includegraphics[width=8cm,height=6cm]{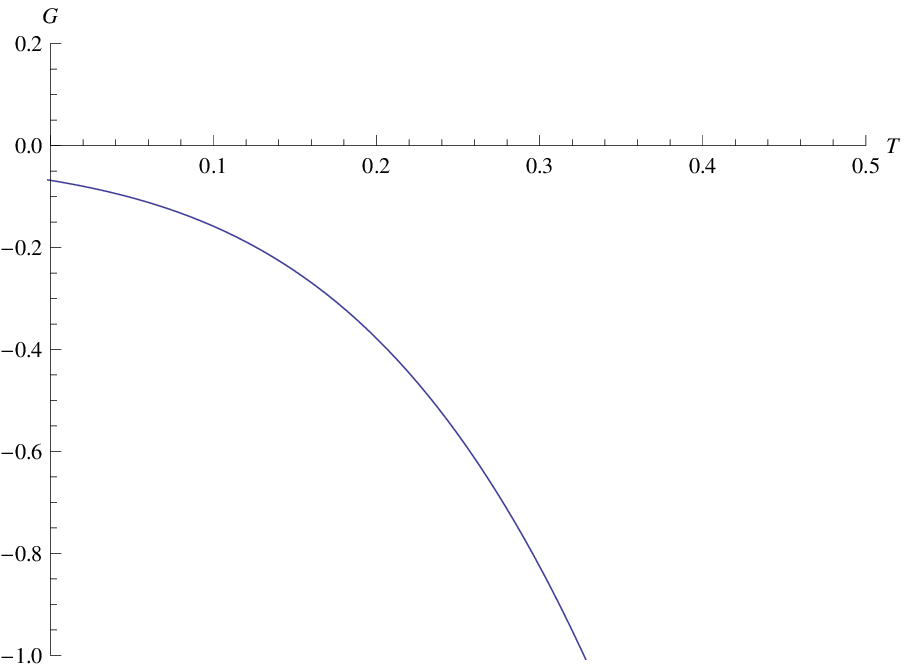}}}
\centerline{\subfigure[]{\label{6c}
\includegraphics[width=8cm,height=6cm]{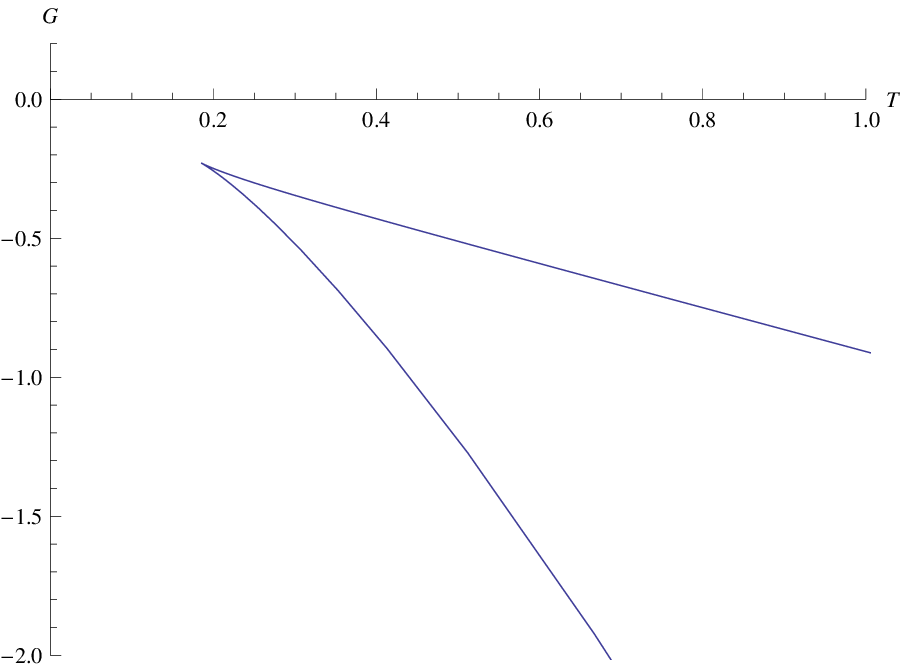}}}
 \caption{$G$ vs. $T$ for
(a) $k=1,\Phi=2$ (b) $k=0,\Phi=2$ (c) $k=-1,\Phi=2$} \label{fg6}
\end{figure*}

 To confirm the phase transition in grand-canonical ensemble, we would like to perform an
    analysis of the behavior of the Gibbs potential. The Gibbs potential which is defined
    by $G=M-TS-\Phi Q$ can be obtained as
 \begin{align}
G=&\frac{\kappa^2\mu^2\Omega_k}{64l^2x_+(k+x_+^2)}\times[2(k+x_+^2)(2k^2+5kx_+^2-x_+^4)\nonumber
\\
&+2kln x_+(2k^2+x_+^2\Phi^2-4kx_+^2-6x_+^4)-x_+^2(2k+x_+^2)\Phi^2]
 .\label{99}
\end{align}
Figure~\ref{fg6} shows the Gibbs potential vs. the temperature for three different cases. For $k=1,0$, the Gibbs
potential decreases with the increasing of the Hawking temperature.
For $k=-1$, two wings are connected at the point where Gibbs
potential reaches a maximum value. According to traditional
thermodynamics, the system is most unstable at this point and would
eventually passes to the locally stable phase which has the low
Gibbs potential corresponding to the lower wing.

\begin{figure*}
\centerline{\subfigure[]{\label{7a}
\includegraphics[width=8cm,height=6cm]{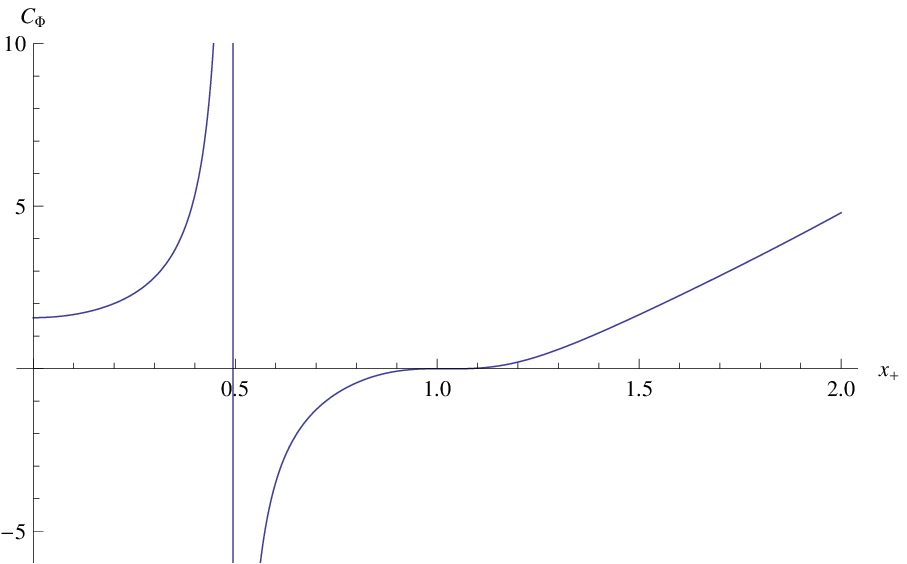}}
\subfigure[]{\label{7b}
\includegraphics[width=8cm,height=6cm]{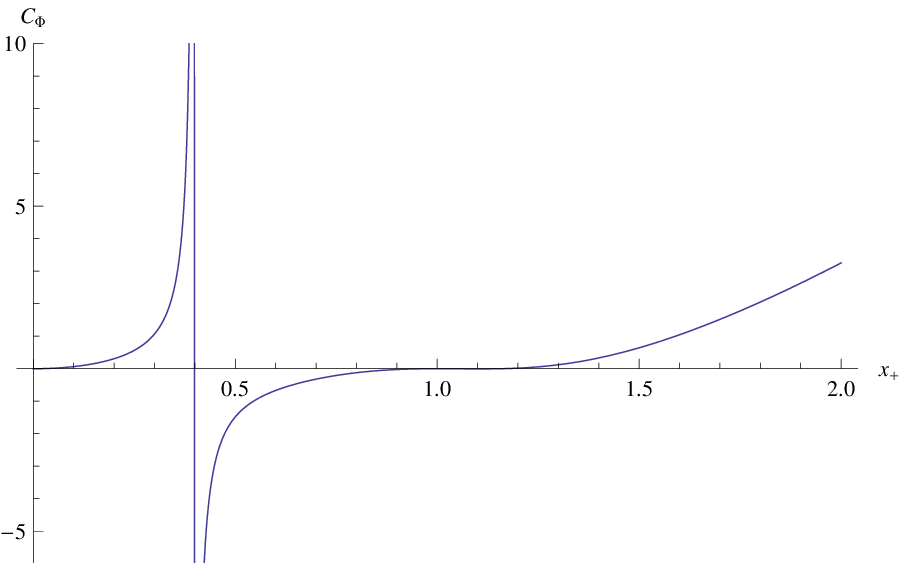}}}
\centerline{\subfigure[]{\label{7c}
\includegraphics[width=8cm,height=6cm]{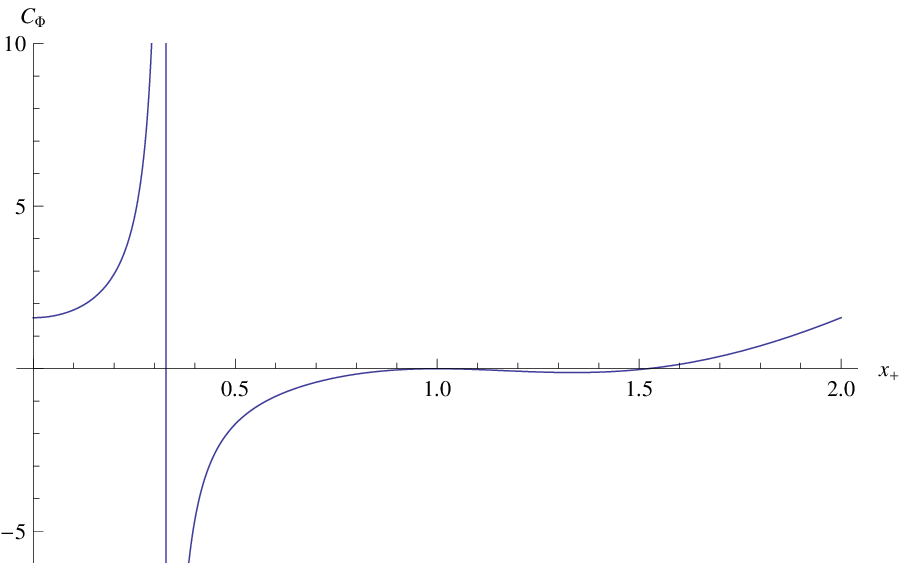}}
\subfigure[]{\label{7d}
\includegraphics[width=8cm,height=6cm]{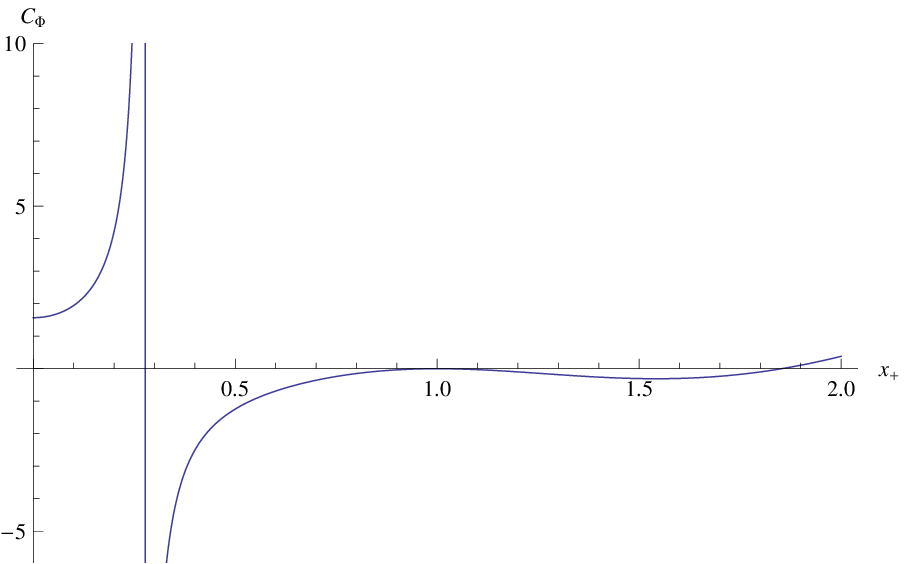}}}
 \caption{$C_\Phi$ vs. $x_+$ for
(a) $k=-1,\Phi=1$ (b) $k=-1,\Phi=2$ (c) $k=-1,\Phi=3$ (d)
$k=-1,\Phi=4$} \label{fg7}
\end{figure*}

\begin{figure*}
\centerline{\subfigure[]{\label{8a}
\includegraphics[width=8cm,height=6cm]{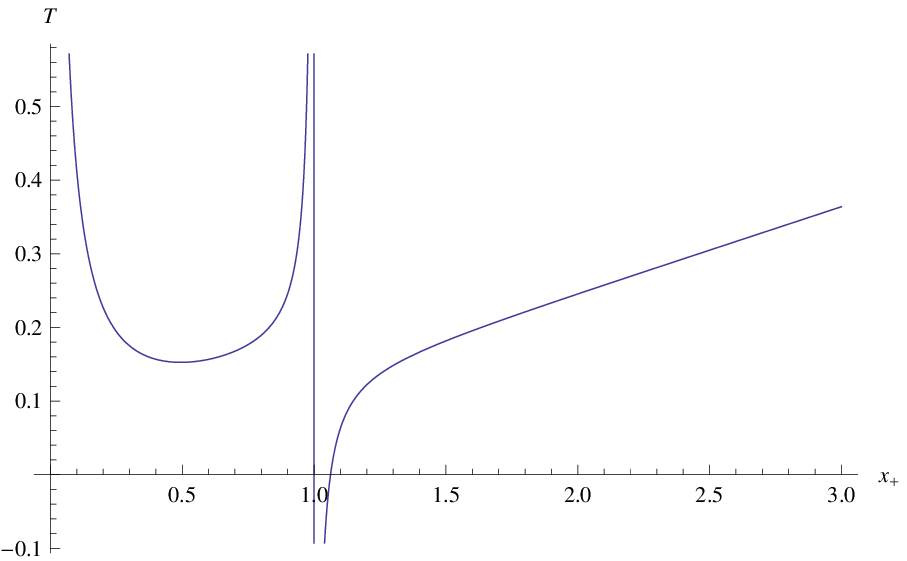}}
\subfigure[]{\label{8b}
\includegraphics[width=8cm,height=6cm]{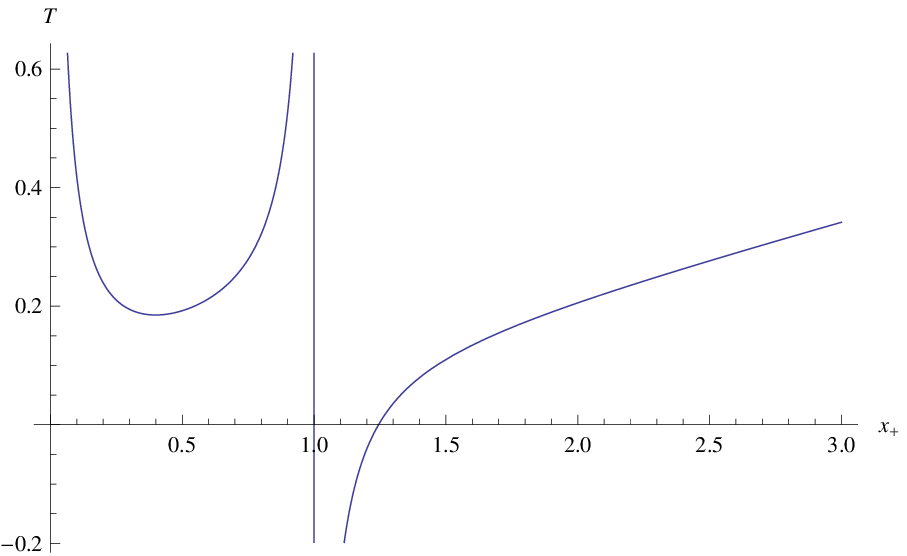}}}
\centerline{\subfigure[]{\label{8c}
\includegraphics[width=8cm,height=6cm]{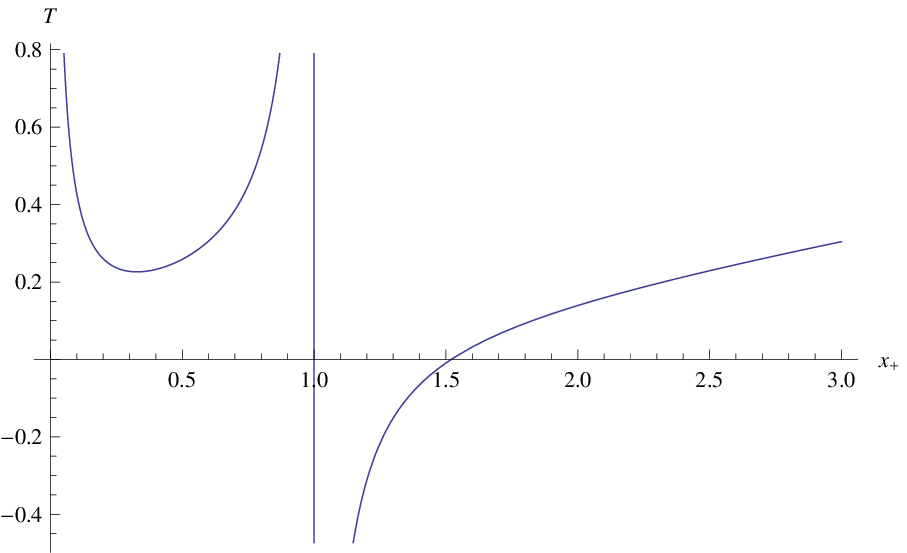}}
\subfigure[]{\label{8d}
\includegraphics[width=8cm,height=6cm]{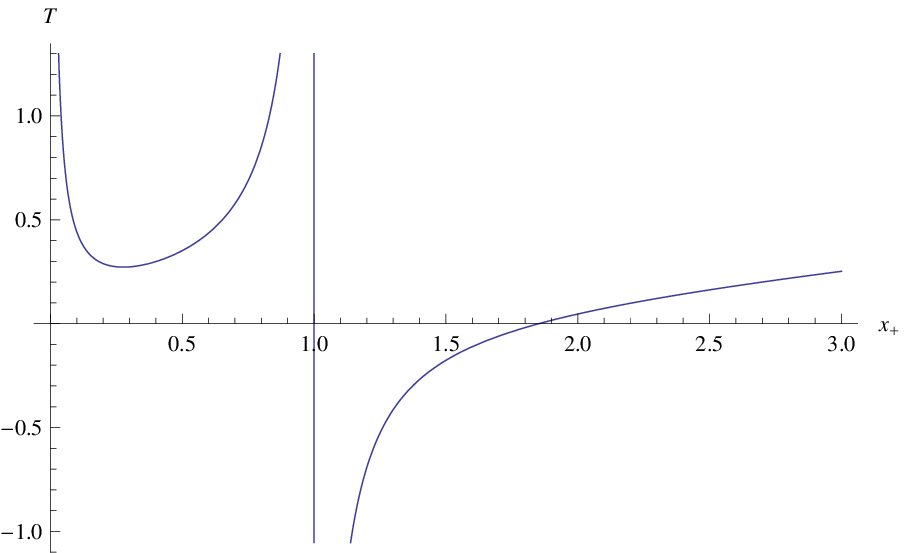}}}
 \caption{$T$ vs. $x_+$ for
(a) $k=-1,\Phi=1$ (b) $k=-1,\Phi=2$ (c) $k=-1,\Phi=3$ (d)
$k=-1,\Phi=4$} \label{fg8}
\end{figure*}

\begin{figure*}
\centerline{\subfigure[]{\label{9a}
\includegraphics[width=8cm,height=6cm]{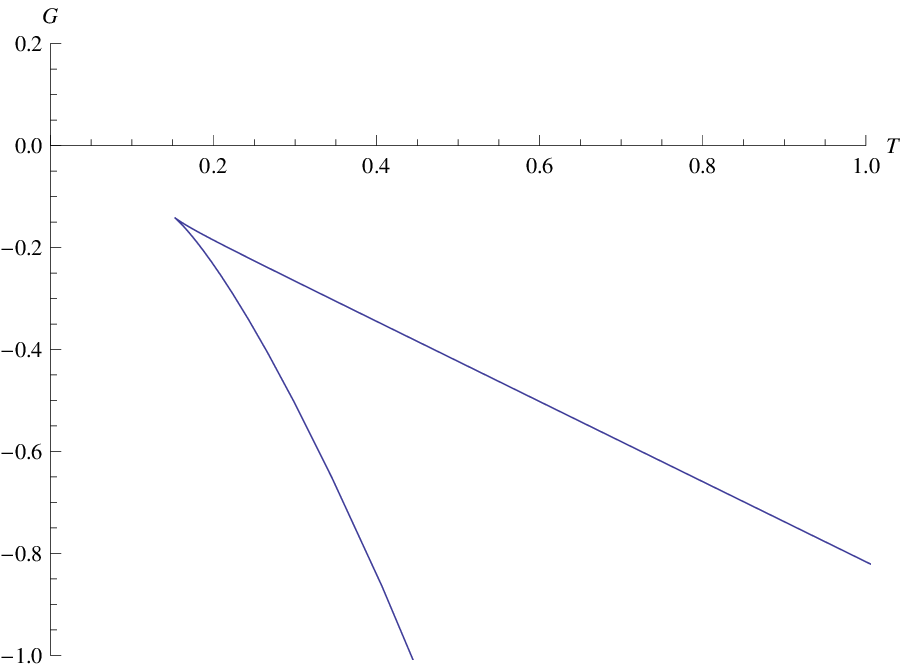}}
\subfigure[]{\label{9b}
\includegraphics[width=8cm,height=6cm]{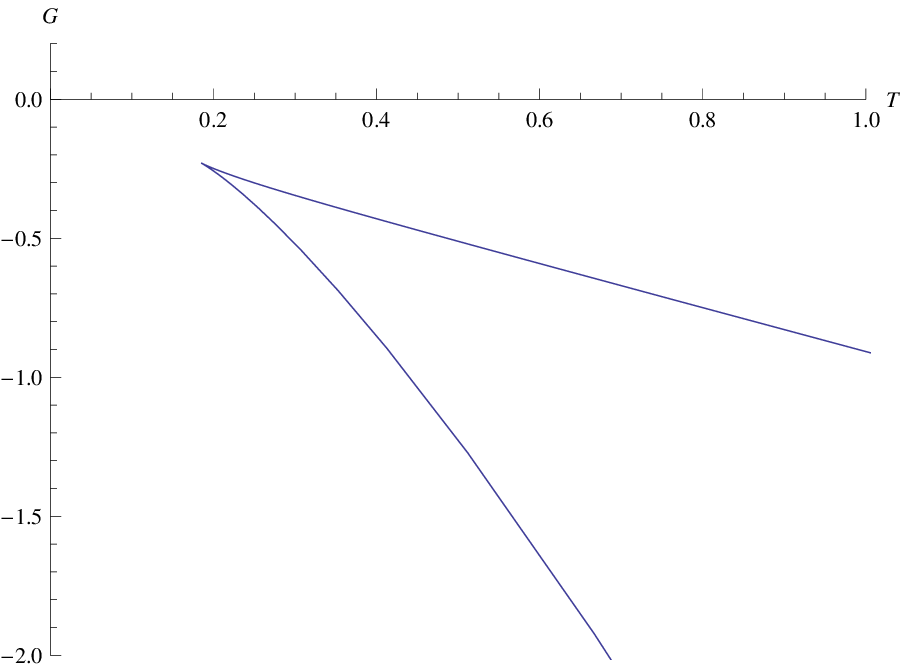}}}
\centerline{\subfigure[]{\label{9c}
\includegraphics[width=8cm,height=6cm]{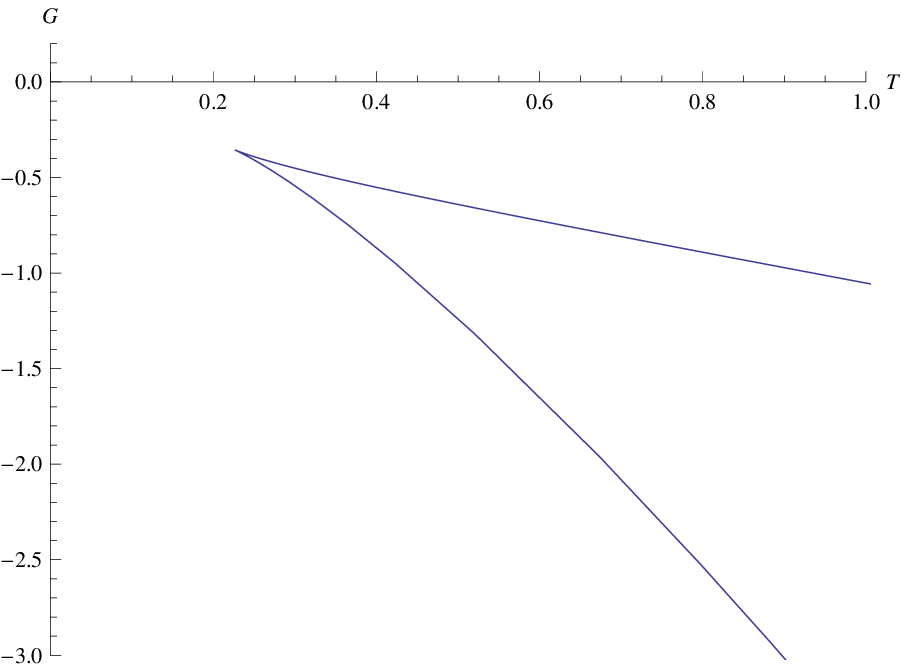}}
\subfigure[]{\label{9d}
\includegraphics[width=8cm,height=6cm]{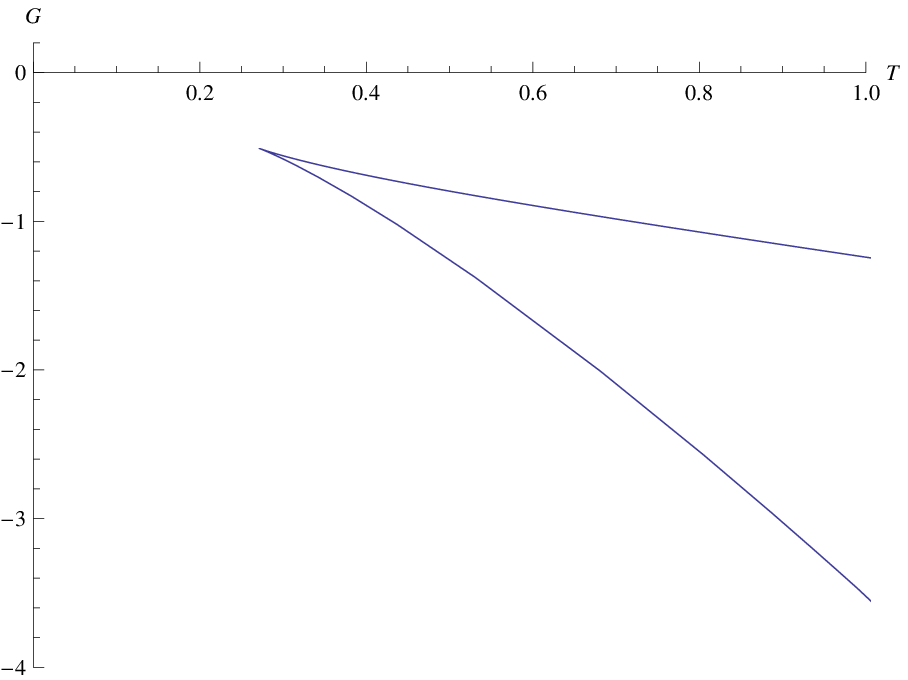}}}
 \caption{$G$ vs. $T$ for
(a) $k=-1,\Phi=1$ (b) $k=-1,\Phi=2$ (c) $k=-1,\Phi=3$ (d)
$k=-1,\Phi=4$} \label{fg9}
\end{figure*}

According to Eq.(\ref{28}), the phase transition point may depend on
the value of the potential $\Phi$. To trace the variation of the
phase transition point due to potential, we plot Figure~\ref{fg7}
with different potential values. In Figure~\ref{fg7}, we can see
that the phase transition point changes with potential $\Phi$. When
$\Phi$ increases, the value of $x_+$ corresponding to the phase
transition point tends to decrease. Figure~\ref{fg8} which shows the
behavior of the Hawking temperature with different choices of
potential values indicates that the phase transition points all
locate in the physical regions. With the increasing of the potential,
the minimum temperature tends to increase while the value of $x_+$
corresponding to the minimum Hawking temperature tends to decrease.
It is quite interesting to note from Figure~\ref{fg9} that the
point having the maximum Gibbs potential also corresponds to the
minimum Hawking temperature. With the increasing of the potential, the
minimum temperature tends to increase, which is consistent with Figure~\ref{fg8}. Finally, we can draw the conclusion
that the local phase transition points not only witness the
divergence of the specific heat but also witness the minimum
temperature and maximum Gibbs potential.

    The phase transition in the grand-canonical ensemble has never been reported before.
    It is similar to the canonical ensemble in which the phase transition only takes place
    for $k=-1$. It is very interesting that the location of the phase transition point changes with potential,
    which is different from canonical ensemble where the phase transition point is independent of the parameters.

\begin{figure*}
\centerline{\subfigure[]{\label{10a}
\includegraphics[width=8cm,height=6cm]{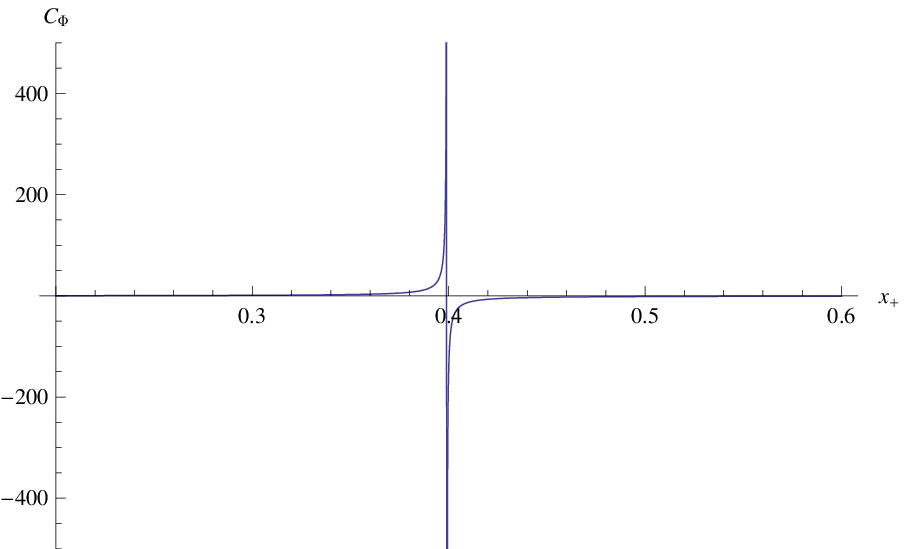}}
\subfigure[]{\label{10b}
\includegraphics[width=8cm,height=6cm]{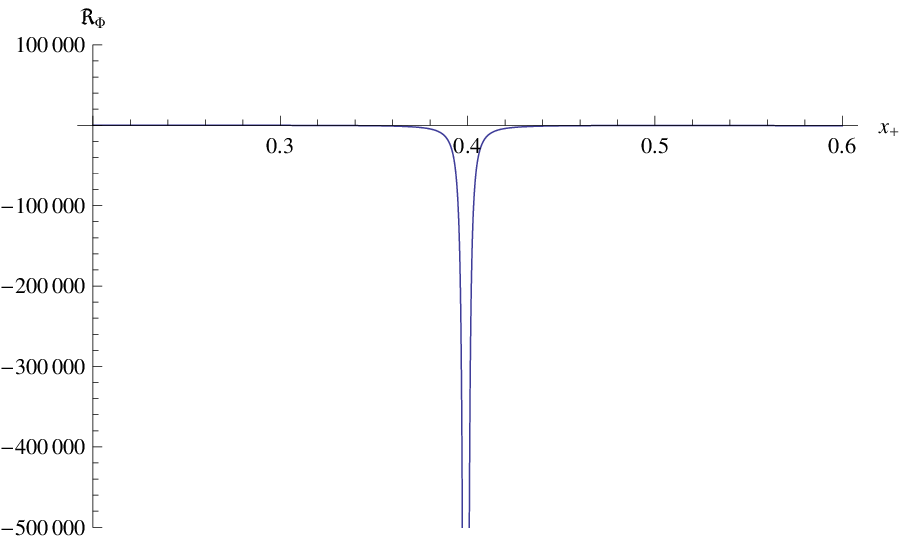}}}
 \caption{($a$) $C_\Phi$ vs. $x_+$ for $k=-1,\Phi=2$ ($b$) $\mathfrak{R}_\Phi$ vs. $x_+$ for
$k=-1,\Phi=2$ } \label{fg10}
\end{figure*}

   To construct geometrothermodynamics in fixed-potential ensemble, we define the thermodynamic potential
   as \cite{{Myung6}}
\begin{equation}
J_\Phi=M-\Phi Q. \label{29}
\end{equation}%
Substituting Eqs.(\ref{5})-(\ref{7}) into Eq.(\ref{29}), we can
obtain the explicit form of $J_\Phi$ as
\begin{equation}
J_\Phi=\frac{\kappa^2\mu^2\Omega_k[2(k+x_+^2)^2-(\Phi^2-\Phi_0^2)x_+^2]}{32l^2x_+}.
\label{30}
\end{equation}%
It is easy to conclude from Eq.(\ref{29}) that
\begin{equation}
dJ_\Phi=dM-(\Phi dQ+Qd\Phi)=TdS-Qd\Phi. \label{31}
\end{equation}%
Unlike the fixed-charge ensemble, $S,\Phi$ are taken as extensive
variables here. And  the corresponding thermodynamic phase space is
a 5-dimensional one coordinated by the set of independent
coordinates\{$J_\Phi,S,\Phi,T,-Q$\}. The fundamental Gibbs 1- form
defined on $\mathcal {T}$ can then be written as
\begin{equation}
\Theta=dJ_\Phi-TdS+Qd\Phi.\label{32}
\end{equation}%
The non-degenerate metric $G$ from Eq.(\ref{11}) can be written as
\begin{equation}
G=(dJ_\Phi-TdS+Qd\Phi)^2+(TS-Q\Phi)(-dSdT-d\Phi dQ).\label{33}
\end{equation}%
Introducing the map
\begin{equation}
\varphi:\{S,\Phi\}\mapsto\{J_\Phi(S,\Phi),S,\Phi,\frac{\partial
J_\Phi}{\partial S},\frac{\partial J_\Phi}{\partial
\Phi}\},\label{34}
\end{equation}%
the space of thermodynamic equilibrium states can be induced.
According to Eq.(\ref{12}), the thermodynamic metric $g$ can be
written as follows
\begin{equation}
g=(S\frac{\partial J_\Phi}{\partial S}+\Phi\frac{\partial
J_\Phi}{\partial \Phi})(-\frac{\partial^2 J_\Phi}{\partial
S^2}dS^2+\frac{\partial^2 J_\Phi}{\partial
\Phi^2}d\Phi^2).\label{35}
\end{equation}%
Utilizing Eqs.(\ref{4}),(\ref{5}) and (\ref{30}), we can easily
calculate the relevant quantities in Eq.(\ref{35})as
\begin{align}
\frac{\partial J_\Phi}{\partial
S}&=\frac{-\Phi^2x_+^2-2(k-3x_+^2)(k+x_+^2)}{16l^2\pi
x_+(k+x_+^2)},\label{36}\\
\frac{\partial J_\Phi}{\partial
\Phi}&=-\frac{\kappa^2\mu^2\Omega_kx_+\Phi}{16l^2},\label{37}\\
\frac{\partial^2 J_\Phi}{\partial
\Phi^2}&=-\frac{\kappa^2\mu^2\Omega_kx_+}{16l^2},\label{38}\\
\frac{\partial^2 J_\Phi}{\partial
S^2}&=\frac{2k^3+10k^2x_+^2+14kx_+^4+6x_+^6-x_+^2(k-x_+^2)\Phi^2}{8l^2\pi^2\kappa^2\mu^2\Omega_kx_+(k+x_+^2)^3}.\label{39}
\end{align}
Comparing Eqs.(\ref{36}) , (\ref{37})with Eqs.(\ref{3}), (\ref{6}), we
can find
\begin{equation}
\frac{\partial J_\Phi}{\partial S}=T,\quad\frac{\partial
J_\Phi}{\partial \Phi}=-Q.\label{40}
\end{equation}%
The results of Eq.(\ref{40}) coincide with Eq.(\ref{31})exactly.
Substituting Eqs.(\ref{36})-(\ref{39}) into Eq.(\ref{35}), we can
find the component of the thermodynamic metric $g$ as
\begin{align}
g_{\Phi\Phi}&=\frac{E(x_+,\Phi)}{1024\pi\kappa^2\mu^2\Omega_kl^4(k+x_+^2)^2},\label{41}\\
g_{SS}&=-\frac{F(x_+,\Phi)}{512\pi^3l^4
\kappa^2\mu^2\Omega_kx_+^2(k+x_+^2)^4},\label{42}
\end{align}%
where
\begin{align}
E(x_+,\Phi)&=4\pi\kappa^4\mu^4\Omega_k^2x_+^2\Phi^2(k+x_+^2)
\nonumber
\\
&-\kappa^2\mu^2\Omega_k[4kx_+^2-2k^2+x_+^2(6x_+^2-\Phi^2)][S_0+\pi
\kappa^2\mu^2\Omega_k(x_+^2+2k\ln x_+)],\label{61}
\\
F(x_+,\Phi)&=[-2k^2+4kx_+^2+x_+^2(6x_+^2-\Phi^2)]\times[4S_0+\pi
\kappa^2\mu^2\Omega_k(x_+^2+2k\ln x_+)]\nonumber
\\
&-4\pi\kappa^2\mu^2\Omega_k\Phi^2x_+^2(k+x_+^2)[2k^3+10k^2x_+^2+kx_+^2(14x_+^2-\Phi^2)+x_+^4(6x_+^2+\Phi^2)].\label{43}
\end{align}
So the Legendre invariant scalar curvature can be given as
\begin{equation}
\mathfrak{R}_\Phi=\frac{W(x_+,\Phi)}{Y(x_+,\Phi)},\label{44}
\end{equation}%
where
\begin{align}
Y(x_+,\Phi)=&\kappa^2\mu^2\Omega_k[2k^3+10k^2x_+^2+kx_+^2(14x_+^2-\Phi^2)+x_+^4(6x_+^2+\Phi^2)]^2
\nonumber
\\
&\times\{4S_0[2k^2-4kx_+^2+x_+^2(\Phi^2-6x_+^2)]+2k\pi
\kappa^2\mu^2\Omega_k[2k^2-4kx_+^2+x_+^2(\Phi^2-6x_+^2)]\ln x_+
\nonumber
\\
&+\pi^2\kappa^2\mu^2\Omega_kx_+^2[2k^2-x_+^2(6x_+^2-5\Phi^2)+4k\Phi^2)]\}^3.\label{45}
\end{align}%
From Eq.(\ref{45}), we can find that the Legendre invariant scalar
curvature may diverge when
$2k^3+10k^2x_+^2+kx_+^2(14x_+^2-\Phi^2)+x_+^4(6x_+^2+\Phi^2)=0$,
which coincides with Eq.(\ref{28}). That is the exact point where
the phase transition may take place. To get an intuitive sense on
this issue, we plot Figure~\ref{fg10}, which shows the correpondence
of the divergence of specific heat $C_\Phi$ and the thermodynamic
scalar curvature of $\mathfrak{R}_\Phi$. From Figure~\ref{fg10}, we
learn that the Legendre invariant metric constructed in
geometrothermodynamics correctly produces the behavior of the
thermodynamic interaction and phase transition structure.

\section{Analytical check of Ehrenfest equations in the fixed-potential ensemble}
\label{Sec5}
\begin{figure*}
\centerline{\subfigure[]{\label{11a}
\includegraphics[width=8cm,height=6cm]{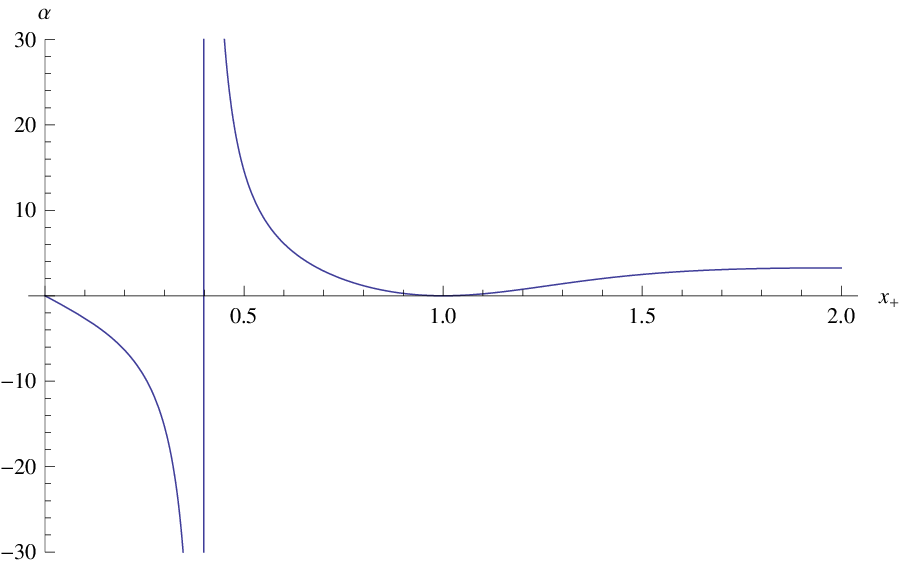}}
\subfigure[]{\label{11b}
\includegraphics[width=8cm,height=6cm]{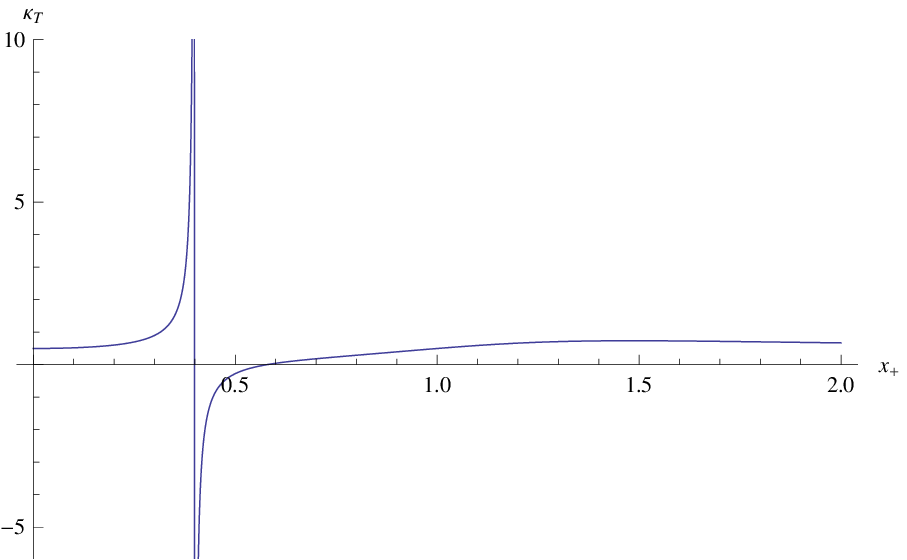}}}
 \caption{ ($a$) $\alpha$ vs. $x_+$ for
$k=-1,\Phi=2$ ($b$) $\kappa_T$ vs. $x_+$ for $k=-1,\Phi=2$ }
\label{fg11}
\end{figure*}

    To investigate the grand-canonical ensemble phase transition, we would like to carry out an analytical check of
     Ehrenfest equations for this case. In conventional thermodynamics, Ehrenfest¡¯s equations are given as
\begin{align}
(\frac{\partial P}{\partial
T})_S&=\frac{C_{P_2}-C_{P_1}}{VT(\alpha_2-\alpha_1)}=\frac{\Delta
C_P}{VT\Delta \alpha},\label{46}\\
(\frac{\partial P}{\partial
T})_V&=\frac{\alpha_2-\alpha_1}{\kappa_{T_2}-\kappa_{T_1}}=\frac{\Delta
\alpha}{\Delta\kappa_T}.\label{47}
\end{align}%
  Taking the similar approach as Ref.\cite{Banerjee5} and considering the analogy ($V\leftrightarrow Q,P\leftrightarrow-\Phi$) between
  the thermodynamic state variables and black hole parameters, we can easily
  write down the corresponding Ehrenfest¡¯s equations for the charged topological
  black hole as
\begin{align}
-(\frac{\partial \Phi}{\partial
T})_S&=\frac{C_{\Phi_2}-C_{\Phi_1}}{TQ(\alpha_2-\alpha_1)}=\frac{\Delta
C_\Phi}{TQ\Delta \alpha},\label{48}\\
-(\frac{\partial \Phi}{\partial
T})_Q&=\frac{\alpha_2-\alpha_1}{\kappa_{T_2}-\kappa_{T_1}}=\frac{\Delta
\alpha}{\Delta\kappa_T},\label{49}
\end{align}%
where $\alpha=\frac{1}{Q}(\frac{\partial Q}{\partial T})_\Phi$ is
the analog of volume expansion coefficient and
$\kappa_T=\frac{1}{Q}(\frac{\partial Q}{\partial \Phi})_T$ is the
analog of isothermal compressibility. Utilizing
Eqs.(\ref{3})-(\ref{6}), we can derive the explicit forms of
relevant quantities as
\begin{align}
\alpha&=\frac{1}{Q}(\frac{\partial Q}{\partial
T})_\Phi=\frac{16l^2\pi
(k+x_+^2)^2x_+}{2k^3-k\Phi^2x_+^2+10k^2x_+^2+\Phi^2x_+^4+14kx_+^4+6x_+^6},\label{50}\\
\kappa_T&=\frac{1}{Q}(\frac{\partial Q}{\partial
\Phi})_T=\frac{(\Phi^2x_+^2+2(k+x_+^2)^2)(k+3x_+^2)}{\Phi(2k^3-k\Phi^2x_+^2+10k^2x_+^2+\Phi^2x_+^4+14kx_+^4+6x_+^6)}.\label{51}
\end{align}%
It is quite interesting to note that $C_\Phi,\alpha,\kappa_T$ share
the same factor in their denominators, namely
$(2k^3-k\Phi^2x_+^2+10k^2x_+^2+\Phi^2x_+^4+14kx_+^4+6x_+^6)$, which
implies that $\alpha,\kappa_T$ may also diverge at the critical
point. To witness the divergence of $\alpha,\kappa_T$, we plot them
in Figure~\ref{fg11}.

Now let's embark on checking the validity of Ehrenfest¡¯s equations
(\ref{48})-(\ref{49}) at the critical point. Note that
\begin{equation}
Q\alpha=(\frac{\partial Q}{\partial T})_\Phi=(\frac{\partial
Q}{\partial S})_\Phi(\frac{\partial S}{\partial
T})_\Phi=(\frac{\partial Q}{\partial S})_\Phi(\frac{C_\Phi
}{T}),\label{52}
\end{equation}%
then the R.H.S of Eq.(\ref{48}) can be transformed into
\begin{equation}
\frac{\Delta C_\Phi}{TQ\Delta \alpha}=[(\frac{\partial S}{\partial
Q})_\Phi]_{cri},\label{53}
\end{equation}%
    where the subscript "cri" denotes the value at the critical point. Utilizing Eqs.(\ref{4})-(\ref{6}) and (\ref{53}), we obtain
\begin{equation}
\frac{\Delta C_\Phi}{TQ\Delta \alpha}=\frac{8\pi
l^2(k+x_{+c}^2)}{\Phi x_{+c}},\label{54}
\end{equation}%
    where $x_{+c}$ denotes the value of $x_+$ at the critical point. The L.H.S of Eq.(\ref{48}) can be derived as
\begin{equation}
-[(\frac{\partial \Phi}{\partial T})_S]_{cri}=\frac{8\pi
l^2(k+x_{+c}^2)}{\Phi x_{+c}}.\label{55}
\end{equation}%
    From Eqs.(\ref{54}) and (\ref{55}), we can draw the conclusion that the first equation of Erhenfest equations is valid at the
    critical point. The L.H.S of Eq.(\ref{49}) can be obtained as
\begin{equation}
-[(\frac{\partial \Phi}{\partial T})_Q]_{cri}=\frac{16\pi l^2\Phi
x_{+c}(k+x_{+c}^2)^2}{(k+3x_{+c}^2)(\Phi^2x_{+c}^2+2(k+x_{+c}^2)^2]}.\label{56}
\end{equation}%
        From the thermodynamic identity \cite{Banerjee5}
\begin{equation}
(\frac{\partial Q}{\partial \Phi})_T(\frac{\partial \Phi}{\partial
T})_Q(\frac{\partial T}{\partial Q})_\Phi=-1,\label{57}
\end{equation}%
we can derive that
\begin{equation}
Q\kappa_T=(\frac{\partial Q}{\partial \Phi})_T=-(\frac{\partial
T}{\partial \Phi})_Q(\frac{\partial Q}{\partial
T})_\Phi=-(\frac{\partial T}{\partial \Phi})_QQ\alpha,\label{58}
\end{equation}%
from which we can calculate the R.H.S of Eq.(\ref{49}) and get
\begin{equation}
\frac{\Delta \alpha}{\Delta \kappa_T}=-[(\frac{\partial
\Phi}{\partial T})_Q]_{cri}=\frac{16\pi l^2\Phi
x_{+c}(k+x_{+c}^2)^2}{(k+3x_{+c}^2)(\Phi^2x_{+c}^2+2(k+x_{+c}^2)^2]}.\label{59}
\end{equation}%
                Eqs.(\ref{56}) and (\ref{59}) reveal the validity of the second equation of Ehrenfest equations. So far,
                we have proved that both the Ehrenfest equations are correct at the critical point. Utilizing Eq.(\ref{28}), we can prove that
                Eqs.(\ref{54})and (\ref{59}) has the same value at the critical point. And the Prigogine-Defay(PD)
                ratio is
\begin{equation}
\Pi=\frac{\Delta C_\Phi \Delta \kappa_T}{T_cQ(\Delta
\alpha)^2}=1.\label{60}
\end{equation}%
                Eq.(\ref{60}) and the validity of Ehrenfest equations show that the grand-canonical ensemble phase transition of the charged topological black hole
                in HL gravity is a second order transition. It is necessary to emphasize that although
                $C_\Phi,\alpha,\kappa_T$ diverge at the critical point, they can cancel each other and allow the R.H.S of  Eqs.(\ref{48})
                and  (\ref{49})  to be finite.
\section{Conclusions}
\label{Sec6}
    The phase transition of a charged topological black hole in Ho\v{r}ava-Lifshitz gravity has been investigated in fixed-charge ensemble and fixed-potential ensemble respectively. To build up geometrothermodynamics in both ensembles, we choose
      the corresponding thermodynamic potential and build up both thermodynamic phase space and the space of thermodynamic equilibrium states.
      In the fixed-charge ensemble, the critical point at which the specific heat $C_Q$ diverges wittnesses the divergence of the corresponding Legendre invariant thermodynamic scalar curvature
      $\mathfrak{R}_Q$. In the fixed-potential ensemble, the critical point at which specific heat $C_\Phi$ diverges also wittnesses the divergence of the corresponding Legendre invariant thermodynamic scalar curvature $\mathfrak{R}_\Phi$. Based on the above results, we can surely conclude that the Legendre invariant metrics constructed
      in geometrothermodynamics can correctly produce the behavior of the thermodynamic interaction and phase transition structure, no matter which
      ensemble is chosen. The Legendre invariant metrics constructed here have successfully predicted the phase
      transition of the charged topological black hole in H-L gravity while other metrics failed~\cite{Cao}, where Cao et al. regarded it
      as \textquotedblleft one exception \textquotedblright.

    The concrete phase transition structures are different due to the choice of ensemble, but both of them can build up geometrothermodynamics. The research in both complementary ensembles
allows us to investigate the phase transition from
    different perspectives and gain a unified picture. We have discovered for the first time the phase transition of a charged topological black hole in Ho\v{r}ava-Lifshitz gravity in the grand-canonical ensemble. It is similar to the canonical ensemble where the phase
    transition only takes place for $k=-1$. However it is found that the location of the phase transition point depends
    on the value of potential, which is different from the canonical ensemble where the phase transition point is
    independent of the parameters. After an analytical check of Ehrenfest scheme, we find that the new phase transition is a
    second order one.

    It is worth noting that the dependence of the thermodynamic scalar
curvature on the ensemble does not contradict the property of
Legendre invariance. Firstly, the phase structures are different due
to different behaviors of the specific heat. This phenomenon should
be attributed to different boundary conditions corresponding to
different choice of ensemble. The stability of black holes turns out
to depend on the choice of boundary conditions and consequently on
the ensemble~\cite{Quevedo110}. Secondly, the motivation of
thermodynamic geometry method is to find the relation between the
divergence of thermodynamic curvature scalar and the existence of
phase transition. So a good thermodynamic metric should reproduce
the behavior of the system no matter which ensemble is chosen. Since
the phase structures are different in different ensembles, we should
not expect the thermodynamic scalar curvature keep invariant.
Recently, the founder of geometrothermodynamics
Quevedo~\cite{Quevedo110} also discussed the ensemble dependence of
geometrothermodynamics. In that paper, the concepts of "total
Legendre transformations" and "partial Legendre transformations"
were put forward. If all the extensive variables change through
Legendre transformation, this kind of Legendre transformation has
been called "total Legendre transformation". Otherwise, if some
extensive variables change through Legendre transformation while
others keep the same, this kind of Legendre transformation has been
called "partial Legendre transformation". The word "partial" means
only some (not all) extensive variables undergo Legendre
transformation. The transformation that relates the potentials
$M(S,Q)$ and $H(S,\phi)$ can serve as an example of "partial
Legendre transformation" because only the extensive variable $Q$ is
transformed into $\phi$ while the the extensive variable $S$ remains
the same during the transformation. The transformation that relates
the potentials $M(S,Q)$ and $G(T,\phi)$ can serve as an example of
"total Legendre transformation" because both of two extensive
variables $S$ and $Q$ are transformed into $T$ and $\phi$
respectively during the transformation. Two types of metrics were
investigated in that paper. It has been proved that a metric which
is invariant under partial Legendre transformation can not be used
to distinguish the thermodynamic properties of different ensembles
naturally. On the contrary, the metric that is only invariant under
a total Legendre transformation can reasonably be used to
distinguish the thermodynamic properties of different ensembles. In
our paper, we have used the metric only invariant under total
Legendre transformation and it has described the black hole
thermodynamics and phase transition successfully.

    In the end, we would like to talk more about the interesting phase transition phenomenon in Ho\v{r}ava-Lifshitz gravity.
    The phase structures in the charged topological black hole are quite different from that in Einstein gravity.
    The phase transition takes place only for $k=-1$ in both complementary ensembles while in Einstein theory, only $k = 1$ case exhibits such
    phase transition. This may be attributed to the ultraviolet behaviour of spacetime in Horava-Lifshitz gravity, as argued by Cao et al.~\cite{Cao}.

\acknowledgments We would like to give great thanks to the anonymous
referee for his helpful suggestions about this paper. This research
is supported by the National Natural Science Foundation of China
(Grant Nos.11235003, 11175019, 11178007). It is also supported by
\textquotedblleft Thousand Hundred Ten\textquotedblright project of
Guangdong Province and Natural Science Foundation of Zhanjiang
Normal University (Grant No. QL1104).


\begin{thebibliography}{99}

\bibitem{Bekenstein}
 J. D. Bekenstein, \emph{Black Holes and Entropy}, \emph{Phys. Rev.} {\bf D 7} (1973) 2333.

\bibitem{Hawking1}
S.W. Hawking, \emph{Particle creation by black holes},\emph{Commum.
Math. Phys.} {\bf 43} (1975) 199.

\bibitem{Hawking2}
 S.W. Hawking and D.N. Page, \emph{Thermodynamics of black holes in anti-de Sitter
 space}, \emph{Comm. Math. Phys.} {\bf 87} (1983) 577.


\bibitem{Chamblin1}
 A. Chamblin, R. Emparan, C.V. Johnson and R.C. Myers, \emph{Charged AdS Black Holes and Catastrophic Holography}, \emph{Phys. Rev.} {\bf D 60} (1999) 064018 [arXiv:hep-th/9902170].


\bibitem{Chamblin2}
A. Chamblin, R. Emparan, C.V. Johnson and R.C. Myers,
\emph{Holography, Thermodynamics and Fluctuations of Charged AdS
Black Holes}, \emph{Phys. Rev.} {\bf D 60} (1999) 104026
[arXiv:hep-th/9904197].


\bibitem{Caldarelli}
 M. M. Caldarelli, G. Cognola and D. Klemm,  \emph{Thermodynamics of Kerr-Newman-AdS Black Holes and Conformal Field Theories}, \emph{Class. Quantum Grav.} {\bf17} (2000) 399-420 [arXiv:hep-th/9908022].


\bibitem{Nojiri}
 S. Nojiri and S. D. Odintsov, \emph{Anti-de Sitter Black Hole Thermodynamics in Higher Derivative Gravity and New Confining-Deconfining Phases in dual CFT}, \emph{Phys. Lett.} {\bf B 521}, (2001) 87-95 [arXiv:hep-th/0109122].


\bibitem{Cai1}
 R. G. Cai, \emph{Gauss-Bonnet Black Holes in AdS Spaces}, \emph{Phys. Rev.} {\bf D 65} (2002) 084014 [arXiv:hep-th/0109133].


\bibitem{Cveti}
     M. Cveti \v{c}, S. Nojiri and S. D. Odintsov, \emph{Black Hole Thermodynamics and Negative Entropy in deSitter and Anti-deSitter Einstein-Gauss-Bonnet gravity}, \emph{Nucl. Phys.} {\bf B 628} (2002) 295-330 [arXiv:hep-th/0112045].


\bibitem{Carlip}
 S. Carlip and S. Vaidya, \emph{Phase Transitions and Critical Behavior for Charged Black Holes}, \emph{Class. Quantum Grav.} {\bf 20} (2003) 3827-3838 [arXiv:gr-qc/0306054].


\bibitem{Cai2}
R. G. Cai and A. Wang, \emph{Thermodynamics and Stability of
Hyperbolic Charged Black Holes}, \emph{Phys. Rev.} {\bf D 70} (2004)
064013 [arXiv:hep-th/0406057].


\bibitem{Myung1}
Y. S. Myung, \emph{No Hawking-Page phase transition in three
dimensions}, \emph{Phys. Lett.} {\bf B 624} (2005) 297
[arXiv:hep-th/0506096].


\bibitem{Carter}
B. M. N. Carter and I. P. Neupane, \emph{Thermodynamics and
Stability of Higher Dimensional Rotating (Kerr) AdS Black Holes},
\emph{Phys. Rev.} {\bf D 72} (2005) 043534 [arXiv:gr-qc/0506103].


\bibitem{Cai3}
 R. G. Cai, S.P. Kim and B. Wang, \emph{Ricci Flat Black Holes and Hawking-Page Phase Transition in Gauss-Bonnet Gravity and Dilaton Gravity}, \emph{Phys. Rev.} {\bf D 76} (2007) 024011 [arXiv:0705.2469v1].


\bibitem{Myung2}
 Y. S. Myung, Y.W. Kim and Y.J. Park, \emph{Thermodynamics and phase transitions in the Born-Infeld-anti-de Sitter black holes}, \emph{Phys. Rev.} {\bf D 78} (2008) 084002 [arXiv:0805.0187v2].


\bibitem{Myung3}
Y. S. Myung, \emph{Phase transition between non-extremal and
extremal Reissner-Nordstrom black holes}, \emph{Mod.Phys.Lett.} {\bf
A 23} (2008) 667-676 [arXiv:0710.2568v3].


\bibitem{Koutsoumbas}
G. Koutsoumbas, E. Papantonopoulos and G. Siopsis, \emph{Phase
Transitions in Charged Topological-AdS Black Hole}, \emph{JHEP} {\bf
0805} (2008) 107 [arXiv:0801.4921v2].


\bibitem{Cadoni}
M. Cadoni, G. D. Appollonio and P. Pani, \emph{Phase transitions
between Reissner-Nordstrom and dilatonic black holes in 4D AdS
spacetime}, \emph{JHEP} {\bf 1003} (2010) 100 [arXiv:0912.3520v3].


\bibitem{Liuhaishan}
 H. S. Liu, H. Lu, M. Luo and K. N. Shao, \emph{Thermodynamical Metrics and Black Hole Phase Transitions}, \emph{JHEP} {\bf 1012} (2010) 054 [arXiv:1008.4482].


\bibitem{Sahay}
A. Sahay, T. Sarkar and G. Sengupta, \emph{On The Phase Structure
and Thermodynamic Geometry of R-Charged Black Holes}, \emph{JHEP}
{\bf 1011} (2010) 125 [arXiv:1009.2236].


\bibitem{Cao}
Q. J. Cao, Y. X. Chen and K. N. Shao, \emph{Black hole phase
transitions in Ho\v{r}ava-Lifshitz gravity}, \emph{Phys. Rev.} {\bf
D 83} (2011) 064015 [arXiv:1010.5044v2].


\bibitem{Quevedo1}
  H. Quevedo, A. Sanchez and S. Taj, \emph{Thermodynamics of topological black holes in Ho\v{r}ava-Lifshitz gravity}, \emph{J. Phys.Phy:Conf.Ser} {\bf 354} (2012) 012015.


\bibitem{Quevedo8}
  H. Quevedo, A. Sanchez, S. Taj and A. Vazquez, \emph{Geometrothermodynamics in Ho\v{r}ava-Lifshitz gravity}, \emph{J. Phys.A-Math.Theor} {\bf 45} (2012) 055211.


\bibitem{Banerjee1}
 R. Banerjee, S. K. Modak and S. Samanta, \emph{Glassy Phase Transition and Stability in Black Holes}, \emph{Eur. Phys. J.} {\bf C 70} (2010) 317 [arXiv:1002.0466].


\bibitem{Banerjee2}
R. Banerjee, S. K. Modak and S. Samanta, \emph{Second Order Phase
Transition and Thermodynamic Geometry in Kerr-AdS Black Hole},
\emph{Phys. Rev.} {\bf D 84} (2011) 064024 [arXiv:1005.4832].


\bibitem{Banerjee3}
 R. Banerjee and D. Roychowdhury, \emph{Critical phenomena in Born-Infeld AdS black holes}, \emph{Phys. Rev.} {\bf D 85} (2011) 044040 [arXiv:1111.0147].


\bibitem{Banerjee4}
 R. Banerjee, S. Ghosh and D. Roychowdhury, \emph{New type of phase transition in Reissner Nordstrom - AdS black hole and its thermodynamic geometry},
    \emph{Phys. Lett.} {\bf B 696}, 156 (2011) 156 [arXiv:1008.2644].


\bibitem{Banerjee5}
 R. Banerjee and D. Roychowdhury, \emph{Thermodynamics of phase transition in higher dimensional AdS black holes}, \emph{JHEP} {\bf 11} (2011) 004 [arXiv:1109.2433].


\bibitem{Banerjee6}
R. Banerjee, S. K. Modak and D. Roychowdhury, \emph{A unified
picture of phase transition: from liquid-vapour systems to AdS black
holes }, \emph{JHEP} {\bf 1210} (2012) 125 [arXiv:1106.3877].


\bibitem{Weishaowen1}
S. W. Wei and Y. X. Liu, \emph{Thermodynamic Geometry of black hole
in the deformed Horava-Lifshitz gravity}, \emph{Europhys.Lett.} {\bf
99} (2012) 20004 [arXiv:1002.1550].


\bibitem{Majhi}
B. R. Majhi and D. Roychowdhury, \emph{Phase transition and scaling
behavior of topological charged black holes in Horava-Lifshitz
gravity}, \emph{Class. Quantum Grav.} {\bf 29} (2012) 245012
[arXiv:1205.0146].


\bibitem{Kim1}
W. Kim and Y. Kim, \emph{Phase transition of quantum corrected
Schwarzschild black hole}, \emph{Phys. Lett.} {\bf B 718} (2012)
687-691 [arXiv:1207.5318].


\bibitem{Tsai}
Y. D. Tsai, X. N. Wu and Y. Yang, \emph{Phase Structure of Kerr-AdS
Black Hole}, \emph{Phys. Rev.} {\bf D 85} (2012) 044005
[arXiv:1104.0502].


\bibitem{Capela}
F. Capela and G. Nardini, \emph{Hairy Black Holes in Massive
Gravity: Thermodynamics and Phase Structure }, \emph{Phys. Rev.}
{\bf D 86} (2012) 024030 [arXiv:1203.4222].


\bibitem{Kubiznak}
D. Kubiznak and  R. B. Mann, \emph{P-V criticality of charged AdS
black holes }, \emph{JHEP} {\bf 1207} (2012)033 [arXiv:1205.0559].


\bibitem{Weishaowen2}
S. W. Wei and Y. X. Liu, \emph{Critical phenomena and thermodynamic
geometry of charged Gauss-Bonnet AdS black holes}, \emph{Phys. Rev.}
{\bf D 87} (2013) 044014 [arXiv:1209.1707].


\bibitem{Eune}
M. Eune, W. Kim and S. H. Yi, \emph{Hawking-Page phase transition in
BTZ black hole revisited }, \emph{JHEP} {\bf 1303} (2013) 020
[arXiv:1301.0395].


\bibitem{Weinhold}
F. Weinhold, \emph{Metric geometry of equilibrium thermodynamics},
\emph{Chem.Phys.} {\bf 63} (1975) 2479.


\bibitem{Ruppeiner}
G. Ruppeiner, \emph{A Riemannian geometric model}, \emph{Phys. Rev.}
{\bf A 20} (1979) 1608.


\bibitem{Salamon}
P. Salamon, E. Ihrig and R. S. Berry, \emph{A group of coordinate
transformations which preserve the metric of Weinhold}, \emph{J.
Math. Phys.} {\bf 24} (1983)2515.


\bibitem{Mrugala}
R. Mrugala, J. D. Nulton, J. C. Schon, and P. Salamon,
\emph{Statistical approach to the geometric structure of
thermodynamics}, \emph{Phys. Rev.} {\bf A 41} (1990) 3156.


\bibitem{Quevedo2}
H. Quevedo, \emph{Geometrothermodynamics}, \emph{J. Math. Phys.}
{\bf 48} (2007) 013506 [arXiv:physics/0604164].


\bibitem{Quevedo3}
H. Quevedo, \emph{Geometrothermodynamics of black holes},
\emph{Gen.Rel.Grav.} {\bf 40} (2008) 971-984 [arXiv:0704.3102].


\bibitem{Quevedo4}
H. Quevedo and A. Sanchez, \emph{Geometrothermodynamics of
asymptotically anti - de Sitter black holes}, \emph{JHEP} {\bf 0809}
(2008) 034 [arXiv:0805.3003].


\bibitem{Alvarez}
J. L. Alvarez, H. Quevedo and A. Sanchez, \emph{Unified geometric
description of black hole thermodynamics}, \emph{Phys. Rev.} {\bf D
77} (2008) 084004 [arXiv:0801.2279].


\bibitem{Quevedo5}
H. Quevedo and A. Sanchez,  \emph{Geometrothermodynamics of black
holes in two dimensions}, \emph{Phys. Rev.} {\bf D 79} (2009) 087504
[arXiv:0902.4488].


\bibitem{Quevedo6}
H. Quevedo and  A. Sanchez, \emph{Geometric description of BTZ black
holes thermodynamics}, \emph{Phys. Rev.} {\bf D 79} (2009) 024012
[arXiv:0811.2524].


\bibitem{Akbar}
M. Akbar, H. Quevedo, K. Saifullah and A. Sanchez, S. Taj,
  \emph{Thermodynamic Geometry Of Charged Rotating BTZ Black Holes},
    \emph{Phys. Rev.} {\bf D 83} (2011) 084031 [arXiv:1101.2722].


\bibitem{Quevedo7}
 H. Quevedo, A. Sanchez, S. Taj and A. Vazquez, \emph{Phase transitions in geometrothermodynamics},
    \emph{Gen.Rel. Grav} {\bf 43} (2011) 1153-1165 [arXiv:1010.5599].


\bibitem{Aviles}
A. Aviles, A. B. Almodovar, L.Campuzano and H. Quevedo,
\emph{Extending the generalized Chaplygin gas model by using
geometrothermodynamics}, \emph{Phys. Rev.} {\bf D 86} (2012) 063508
[arXiv:1203.4637].


\bibitem{Hanyiwen}
Y. W. Han and G. Chen, \emph{Thermodynamics, geometrothermodynamics
and critical behavior of (2+1)-dimensional black hole}, \emph{Phys.
Lett.} {\bf B 714} (2012) 127-130 [arXiv:1207.5626].


\bibitem{Horava1}
P. Hor\v{r}ava, \emph{Quantum Gravity at a Lifshitz Point},
\emph{Phys. Rev.} {\bf D 79} (2009) 084008 [arXiv:0901.3775].


\bibitem{Horava2}
P. Hor\v{r}ava, \emph{Membranes at Quantum Criticality}, \emph{JHEP}
{\bf 0903} (2009) 020 [arXiv:0812.4287].


\bibitem{Horava3}
P. Hor\v{r}ava, \emph{Spectral Dimension of the Universe in Quantum
Gravity at a Lifshitz Point}, \emph{Phys. Rev. Lett} {\bf 102}
(2009) 161301 [arXiv:0902.3657].


\bibitem{Luhong}
H. Lu, J. Mei and C. N. Pope, \emph{Solutions to Horava Gravity},
\emph{Phys. Rev. Lett} {\bf 103} (2009) 091301 [arXiv:0904.1595].


\bibitem{Cai4}
R. G. Cai, L. M. Cao and N. Ohta, \emph{Topological Black Holes in
Horava-Lifshitz Gravity}, \emph{Phys. Rev.} {\bf D 80} (2009) 024003
[arXiv:0904.3670].


\bibitem{Colgain}
E. O. Colgain and H.Yavartanoo, \emph{Dyonic solution of
Horava-Lifshitz Gravity}, \emph{JHEP} {\bf 0908} (2009) 021
[arXiv:0904.4357].


\bibitem{Kehagias}
A.Kehagias and K.Sfetsos, \emph{The black hole and FRW geometries of
non-relativistic gravity}, \emph{Phys. Lett.} {\bf B 678} (2009)
123-126 [arXiv:0905.0477].


\bibitem{Kiritsis}
E. Kiritsis, \emph{Spherically symmetric solutions in modified
Horava-Lifshitz gravity}, \emph{Phys. Rev.} {\bf D 81} (2010) 044009
[arXiv:0911.3164].


\bibitem{Cai5}
R. G. Cai, L. M. Cao and N.Ohta, \emph{Thermodynamics of Black Holes
in Horava-Lifshitz Gravity}, \emph{Phys. Lett.} {\bf B 679} (2009)
504-509 [arXiv:0905.0751].


\bibitem{Cai6}
R. G. Cai and N. Ohta, \emph{Horizon Thermodynamics and
Gravitational Field Equations in Horava-Lifshitz Gravity},
    \emph{Phys. Rev.} {\bf D 81} (2010) 084061 [arXiv:0910.2307].


\bibitem{Chendeyou}
D. Y. Chen, H. T. Yang and X. T. Zu, \emph{Hawking radiation of
black holes in the $z = 4$ Horava-Lifshitz gravity}, \emph{Phys.
Lett.} {bf B 681} (2009) 463 [arXiv:0910.4821].


\bibitem{Gao}
X. Gao, Y. Wang, R. Brandenberger and A. Riotto, \emph{Fluctuations
in a Ho\v{r}ava-Lifshitz Bouncing Cosmology }, \emph{Phys. Rev.}
{\bf D 81} (2010) 083508 [arXiv:0911.3196].


\bibitem{Castillo}
A. Castillo and A. Larranaga, \emph{Entropy for Black Holes in the
Deformed Ho\v{r}ava-Lifshitz Gravity}, \emph{Electron. J. Theor.
Phys.} {\bf 8} (2011) 1-10 [arXiv:0906.4380].


\bibitem{Majhi2}
B. R. Majhi, \emph{Hawking radiation and black hole spectroscopy in
Horava-Lifshitz gravity}, \emph{Phys. Lett.} {\bf B 686} (2010) 49
[arXiv:0911.3239].


\bibitem{Pengjinjun}
J. J. Peng and S. Q. Wu, \emph{Hawking Radiation of Black Holes in
Infrared Modified Ho\v{r}ava-Lifshitz Gravity}, \emph{Eur. Phys. J.}
{\bf C 66} (2010) 325-331 [arXiv:0906.5121].


\bibitem{Myung4}
Y. S. Myung, \emph{Entropy of black holes in the deformed
Ho\v{r}ava-Lifshitz gravity}, \emph{Phys. Lett.} {\bf B 684} (2010)
158 [arXiv:0908.4132].


\bibitem{Myung5}
Y. S. Myung and Y. W. Kim, \emph{Thermodynamics of
Ho\v{r}ava-Lifshitz black holes}, \emph{Eur. Phys. J.} {\bf C 68}
(2010) 265-270 [arXiv:0905.0179].


\bibitem{Myung6}
Y. S. Myung, Y. W. Kim and Y. J. Park, \emph{Ruppeiner geometry and
2D dilaton gravity in the thermodynamics of black holes},
\emph{Phys.Lett.} {\bf B 663} (2008) 342-350 [arXiv:0802.2152].


\bibitem{Quevedo110}
H. Quevedo, A. Sanchez and S. Taj, \emph{On the ensemble dependence
in black hole geometrothermodynamics}, [arXiv:1304.3954v1].

\end{thebibliography}
\end{document}